\documentclass[fleqn,usenatbib]{mnras}

\usepackage[T1]{fontenc}

\DeclareRobustCommand{\VAN}[3]{#2}
\let\VANthebibliography\thebibliography
\def\thebibliography{\DeclareRobustCommand{\VAN}[3]{##3}\VANthebibliography}

\usepackage{graphicx}	
\usepackage{amsmath}	
\usepackage{multicol}   
\usepackage{bm}	    	
\usepackage{pdflscape}	

\usepackage{newtxtext,newtxmath}
\usepackage{enumerate}

\graphicspath{{./}{figures/}}

\title[Dusty/Dust-free Gas Structures of the AGN]{Updated Picture of the Active Galactic Nuclei with Dusty/Dust-free Gas Structures and Effects of the Radiation Pressure}

\author[S. Mizukoshi et al.]{
Shoichiro Mizukoshi,$^{1}$\thanks{E-mail: s.mizukoshi@ioa.s.u-tokyo.ac.jp}
Takeo Minezaki,$^{1}$
Hiroaki Sameshima,$^{1}$
Mitsuru Kokubo,$^{2}$
\newauthor
Hirofumi Noda,$^{3}$
Taiki Kawamuro,$^{4}$
Satoshi Yamada,$^{4}$
and Takashi Horiuchi,$^{1}$
\\
$^{1}$Institute of Astronomy, Graduate School of Science, The University of Tokyo, 2-21-1 Osawa, Mitaka, Tokyo 181-0015, Japan\\
$^{2}$National Astronomical Observatory of Japan, National Institutes of Natural Sciences, 2-21-1 Osawa, Mitaka, Tokyo 181-8588, Japan\\
$^{3}$Astronomical Institute, Tohoku University, 6-3 Aramakiazaaoba, Aoba-ku, Sendai, Miyagi 980-8578, Japan\\
$^{4}$RIKEN Cluster for Pioneering Research, 2-1 Hirosawa, Wako, Saitama 351-0198, Japan
}

\date{Accepted XXX. Received YYY; in original form ZZZ}

\pubyear{2021}

\begin{document}
\label{firstpage}
\pagerange{\pageref{firstpage}--\pageref{lastpage}}
\maketitle

\begin{abstract}
This study investigates the properties of two gas structures of X-ray selected active galactic nuclei (AGNs), that is, dusty and dust-free gas components, by separating them with the line-of-sight dust extinction ($A_V$) and the neutral gas column density ($N_{\mathrm{H}}$). 
The typical column density of the dusty and dust-free gas differs depending on the Seyfert type, indicating that both structures have anisotropic column density distributions. 
The number of targets with the dusty gas column density ($N_{\mathrm{H,d}}$) of $\log N_{\mathrm{H,d}}\ [\mathrm{cm^{-2}}]>23$ is much smaller than that with the same column density of the dust-free gas. 
This result indicates that the optically-thick part of the dusty gas structure is very thin. 
There are very few targets with a larger Eddington ratio ($f_{\mathrm{Edd}}$) than the effective Eddington limit of the dusty gas and the covering factor of the dusty gas with $22\leq \log N_{\mathrm{H,d}}\ [\mathrm{cm^{-2}}]<24$ exhibits a clear drop at the effective Eddington limit. 
These results support the scenario wherein the covering factor of the dusty torus decreases in a high Eddington ratio owing to the radiation-driven dusty gas outflow. The covering factor of the dust-free gas with the column density ($N_{\mathrm{H,df}}$) of $22\leq \log N_{\mathrm{H,df}}\ [\mathrm{cm^{-2}}]<24$ similarly exhibits the decrease in high Eddington ratio, although it may be owing to the dust-free gas outflow driven by certain other mechanisms than the radiation pressure. 
Finally, we propose an updated picture of the AGN gas structure based on our results and the literature.
\end{abstract}




\section{Introduction}
\label{sec:intro}

The structure of active galactic nuclei (AGNs) has been understood based on the unified model \citep{Antonucci85,Antonucci93,Urry95}.
In this model, the supermassive black hole (SMBH) and the accretion disk are at the centre, and they are surrounded by an obscuring structure of dusty material, which is called the dusty torus.
This model interprets the existence or absence of optical broad emission lines, that is, Seyfert type, as simply the difference in a viewing angle.
When we observe the AGN from the face-on side, the optical broad emission lines emitted from the broad-line region (BLR) at the centre are visible; however, they cannot be observed from the edge-on side owing to the attenuation by the dusty torus.
This scenario is supported by the observation of the polarised broad emission lines \citep[e.g.][]{Antonucci85} or broad emission lines in the near-infrared band \citep[e.g.][]{Reunanen03,Smajic12} for type-2 AGNs.

The dusty torus may cause the outflow owing to the strong radiation pressure from the central engine and it may inject the energy into the interstellar medium of the host galaxy and suppress its star formation \citep[][for review]{Fabian12}.
Further, this dusty gas outflow may partly contribute to the polar dust emission, which is observed in the central compact region of the AGN via mid-infrared (MIR) interferometry \citep[e.g.][]{Raban09,Honig12,Honig13,Tristram14,Leftley18,Isbell22}, or in the circumnuclear region via MIR imaging \citep[e.g.][]{Asmus16,Asmus19}.

To investigate the nature of the dusty gas outflow, the amount of dust in the dusty torus must be understood and it is often characterised by the line-of-sight dust extinction.
The line-of-sight $V$-band dust extinction magnitude ($A_V$) is often estimated by the decrement of the flux ratio of two emission lines in the optical band \citep[e.g.][]{Baker38,Ward87,Gaskell17} or in the near-infrared (NIR) band \citep[e.g.][]{Ward87,Maiolino01,Schnorr-Muller16,FRicci22}.
\cite{Shimizu18}  estimated $A_V$ of unobscured AGNs using the ratio of the broad H$\alpha$ and the 14-150 keV X-ray luminosities.
\cite{Burtscher16} used NIR spectroscopic data for the estimation of $A_V$ by focusing on the decrement in the $K$-band colour temperature and measured a relatively large dust extinction of $A_V\lesssim30$ mag.

\cite{Mizukoshi22} estimated $A_V$ by focusing on the reddening of the NIR time-variable flux components.
Although unobscured AGNs exhibit a relatively constant colour of time-variable flux components in the optical/NIR band, it becomes redder in obscured AGNs \citep{Winkler92,Glass04}; thus, we can convert it to the dust extinction.
This method facilitates the measurement of very large dust extinction of $A_V\lesssim65$ mag.
\cite{Mizukoshi22} compared the derived $A_V$ and the line-of-sight neutral gas column density ($N_{\mathrm{H}}$) that was estimated via the hard X-ray observation \citep{Ricci17b}, and reported certain characteristic behaviours: (i) $N_{\mathrm{H}}$ of dust-obscured AGNs is typically larger than that estimated with $A_V$ and the gas-to-dust ratio of the Galactic interstellar medium (ISM), (ii) $N_{\mathrm{H}}$ of each target scatters within more than two orders of magnitude, and (iii) the lower edge of the $N_{\mathrm{H}}$ distribution of obscured AGNs in the $A_V$--$N_{\mathrm{H}}$ diagram is comparable to the typical relation of the Galactic ISM.
These behaviours have been reported in the literature \citep{Burtscher16,Shimizu18}.
To interpret these behaviours, \cite{Mizukoshi22} adopted the scenario wherein the dust-free gas component exists inside the dusty torus.
In this scenario, the dusty torus has the same dust-to-gas ratio as the Galactic value, and the $N_{\mathrm{H}}$ excess from the Galactic value is attributed to the dust-free gas.
This scenario also facilitates an explanation of  the time variation of $N_{\mathrm{H}}$ in months-to-years time scale \citep[][and citation therein]{Burtscher16}.
While the presence of this dust-free gas has been suggested in many other previous studies \citep{Granato97,Burtscher16,Ichikawa19,Ogawa21,Esparza-Arredondo21}, the structure of the dust-free gas region remains unclear.

According to the literature \citep{Fabian06,Fabian08}, the dusty gas has an effective cross section that is considerably larger than the Thomson cross section, hence the dusty gas is blown out at a luminosity much smaller than the Eddington limit.
This luminosity is called the effective Eddington limit ($L_{\mathrm{Edd}}^{\mathrm{eff}}$).
In larger $N_{\mathrm{H}}$, most of the ionizing photon is absorbed by a small fraction of the gas and most of the gas acts as a 'dead weight', then the effective cross section becomes smaller.
The effective cross section is also affected by the amount of dust in the dusty gas \citep{Fabian09}, and it increases if the gas contains more dust.
Some previous studies have compared $N_{\mathrm{H}}$ and the Eddington ratio ($f_{\mathrm{Edd}}$), which is the ratio of the bolometric luminosity ($L_{\mathrm{bol}}$) to the Eddington limit of the AGN, of nearby AGNs \citep[e.g.][]{Fabian08,Fabian09,Ricci17c,Jun21,Ricci22} and showed that there are few targets in the region where $\log N_{\mathrm{H}}\ [\mathrm{cm^{^2}}]\gtrsim22$ and $L_{\mathrm{bol}}>L_{\mathrm{Edd}}^{\mathrm{eff}}$.
This region is called the 'forbidden region'.
They concluded that, if the AGN enters the forbidden region, the dusty gas outflow will occur and the surrounding material will be blown out, and thus become an unobscured AGN in a short time scale. 

Although the forbidden region is defined based on the Eddington ratio and $N_{\mathrm{H}}$ in previous studies, this prescription is somehow inadequate if  the dust-free gas truly exists. 
This is because, in this case, $N_{\mathrm{H}}$ is attributed to both the dusty and dust-free gas.
Hence, using $A_V$, which reflects only the dusty gas, is more applicable instead of $N_{\mathrm{H}}$ to consider the dusty gas outflow accurately.
In this study, we investigate the properties of the structure of both the dusty and dust-free gas by distinguishing these two gas components using $A_V$ and $N_{\mathrm{H}}$.
We also explore the relationship of these gas structures with the Eddington ratio and examine the effects of the radiation pressure on these gas structures.
Based on these results, we finally propose an update of the AGN picture originally proposed by \cite{Ricci17c}.
The remainder of this paper is organized as follows. 
In Sec. \ref{sec:data analysis}, we explain the details of the data and the sample selection.
Sec. \ref{sec:method} describes the measurement of $A_V$ in this study and the separation of the dusty and dust-free gas components.
The results on the properties of these two gas components and  comparison with the Eddington ratio are presented in Sec. \ref{sec:results}.
In Sec. \ref{sec:discussion}, we discuss the properties of the dust-free gas outflow in relation to the properties of the warm absorber \citep[][for review]{Laha21,Gallo23}, and show the updated AGN picture based on the results of this study.
The findings are summarised in Sec.\ref{sec:conclusion}.
Throughout this study, we adopt the cosmology $H_0=70\ \mathrm{km\ s^{-1}\ Mpc^{-1}}$, $\Omega_0=0.30$, and $\Omega_{\Lambda}=0.70.$

\section{data and targets}
\label{sec:data analysis}


\subsection{Details about the data}

The BAT AGN Spectroscopic Survey DR1 \citep[BASS DR1,][]{Koss17,Ricci17b} catalogues 836 nearby AGNs originally provided in the \textit{Swift}/BAT 70-month catalogue \citep{Baumgartner13}.
The BASS DR1 catalogue provides X-ray observational data such as the hard X-ray luminosity and $N_{\mathrm{H}}$ for almost all targets.
It also contains the Seyfert-type and observational data of certain broad optical lines for hundreds of targets, which were obtained by follow-up optical observations or archival data.
In this study, we used the data of $N_{\mathrm{H}}$, the intrinsic 14--150 keV luminosity ($L_{\mathrm{14-150\,keV,intr}}$), and the broad $\mathrm{H\alpha}$ flux ($f_{\mathrm{bH\alpha}}$) of each target if it was available.
The data of $L_{\mathrm{14-150\,keV,intr}}$ and $f_{\mathrm{bH\alpha}}$ were used to measure $A_V$ of unobscured AGNs based on the method of \cite{Shimizu18} (Sec. \ref{subsec:Av calculation} presents further details).

The BASS DR2 catalogue was recently released \citep{Koss22a,Koss22b}.
This catalogue includes the data of $f_{\mathrm{Edd}}$ for most targets.
For the calculation of $f_{\mathrm{Edd}}$, they derived $L_{\mathrm{bol}}$ from $L_{\mathrm{14-150\,keV,intr}}$ in BASS DR1 considering the bolometric correction of eight \citep{Koss22b}.
They also derived the BH mass based on several methods, such as direct (reverberation mapping, OH megamaser, high-quality gas or stellar kinematics) measurements, broad emission lines \citep{Mejia-Restrepo22} for mainly Sy1 AGNs, and stellar velocity dispersion for all Sy1.9 and Sy2 AGNs \citep{Koss22b}.
According to \cite{Koss22b}, the uncertainty of the BH mass based on the stellar velocity dispersion is of order 0.5 dex; hence, $f_{\mathrm{Edd}}$ of Sy1.9, Sy2 AGNs is also considered to be approximately 0.5 dex.
BASS DR2 also contains Seyfert-type data (Sy1, Sy1.9, or Sy2) and certain AGNs have different types from BASS DR1 owing to the update of optical spectroscopic data in BASS DR2. 
Herein, we used the data of $f_{\mathrm{Edd}}$ and  Seyfert type from BASS DR2.

We used the data in $W1$ ($3.4\,\mu$m) and $W2$ ($4.6\,\mu$m) bands of \textit{Wide-field Infrared Survey Explorer} \citep[\textit{WISE},][]{Wright10} as in \cite{Mizukoshi22} to measure $A_V$ of each target (Sec. \ref{subsec:Av calculation}).
The \textit{WISE} mission performed cryogenic all-sky observation in $W1$, $W2$, $W3$ ($11\,\mu$m), and $W4$ ($22\,\mu$m) bands for approximately a year after its launch, and it has performed all-sky monitoring observation in $W1$ and $W2$ bands for approximately 10 years after its reactivation.
This monitoring data is obtained generally once per six months, and the observation is performed for approximately 2--3 days in one observational epoch.
Here, we used the latest version of \textit{WISE} data available as of February 2023.

\begin{figure*}
    \includegraphics[width=\linewidth]{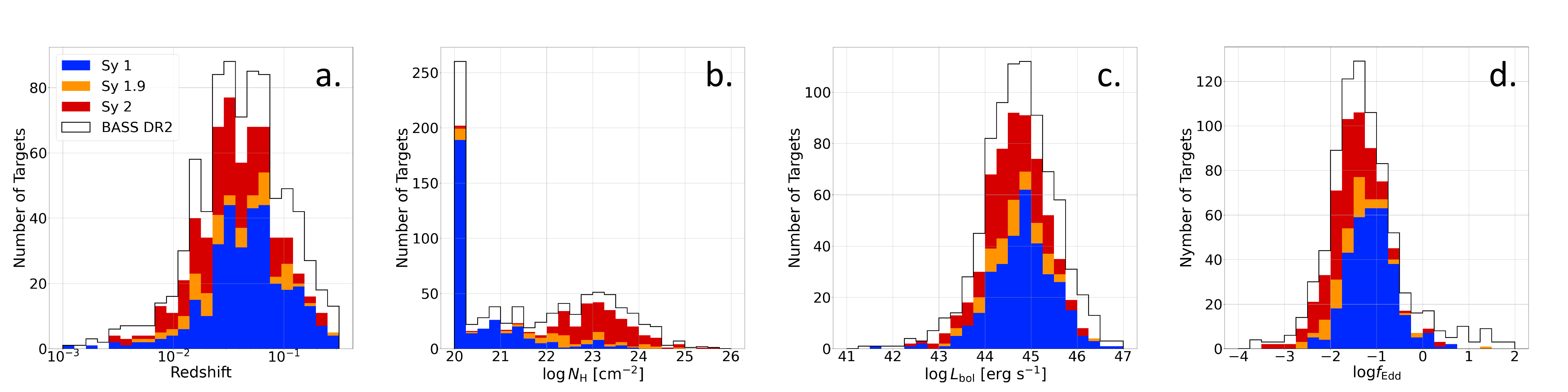}\par 
\caption{The histograms of (a) the redshift, (b) the total gas column density $N_{\mathrm{H}}$, (c) the bolometric luminosity $L_{\mathrm{bol}}$, and (d) the Eddington ratio $f_{\mathrm{Edd}}$ of our sample. The blue, orange, and red histograms show the distribution of Sy1, Sy1.9, and Sy2 AGNs in our sample, respectively. The black line shows the distribution of the complete BASS DR2 sample with $N_{\mathrm{H}}$ or $f_{\mathrm{Edd}}$ data.}
 \label{fig:BASS sample}
\end{figure*}

\subsection{Sample selection}
\label{subsec:BASS sample}
We selected our sample from BASS DR2 catalogue, which comprises 858 nearby AGNs.
The sample selection follows the same criteria as \cite{Mizukoshi22}; (i) non-blazar AGNs, (ii) the redshift is smaller than 0.5, (iii) the Seyfert type is confirmed, (vi) the dust extinction owing to the Galactic ISM is less than 2 mag, (v) the \textit{WISE} data of the target satisfies the criteria for four quality flags of the data \citep[][for detail]{Mizukoshi22}, and (vi) the saturated pixels are less than 10\% of the total pixels for each data in all epochs.
Consequently, 194 targets were eliminated based on these criteria.

In the measurement of $A_V$ based on \cite{Mizukoshi22}, we used the typical ratio of the flux variation amplitude in $W1$- to $W2$-band monitoring data, which is called the flux variation gradient \citep[FVG,][]{Winkler92}.
The FVG can be measured via a regression analysis on the flux-flux plot, wherein we took the flux data of $W1$ and $W2$ bands in the same epoch on the horizontal and vertical axes, respectively \citep{Mizukoshi22}.
Thus, we eliminated 58 samples with weak correlation between the $W1$- and $W2$-band monitoring data, wherein the correlation coefficient was smaller than 0.7, to select targets with accurate FVGs.
After this elimination, we also eliminated four samples with the uncertainty of the derived FVG larger than 0.2.

Finally, we eliminated samples without the data of either $N_{\mathrm{H}}$ in BASS DR1 or $f_{\mathrm{Edd}}$ in BASS DR2.
Our final sample comprised 589 X-ray selected AGNs with 318, 68, and 203 Sy1s, Sy1.9s, and Sy2s, respectively, which is more than 100 targets larger than that of \cite{Mizukoshi22}.
Figure \ref{fig:BASS sample} shows the histograms of the redshift, $N_{\mathrm{H}}$, $L_{\mathrm{bol}}$, and $f_{\mathrm{Edd}}$ of the complete BASS DR2 sample and the final sample of this study.
It shows that our sample selection largely holds the original distributions of the BASS DR2 catalogue.

\section{methods}
\label{sec:method}

\subsection{Calculation of the Dust extinction}
\label{subsec:Av calculation}

We first calculated $A_V$ with \textit{WISE} monitoring data in the same manner as that of \cite{Mizukoshi22}:

\begin{equation}
    A_{V,\mathrm{M22}}=-\frac{5}{2(k_{W1}-k_{W2})}\log \left(\frac{\beta}{\beta_0}\right),
\end{equation}
where $k_{W1}=0.064$ and $k_{W2}=0.045$ are the ratios of the dust extinction in $W1$ ($A_{W1}$) or $W2$ band ($A_{W2}$) to $A_V$, that is, $A_{W1}/A_V$ and $A_{W2}/A_V$, respectively, $\beta$ is the FVG of each target, and $\beta_0=0.86\pm0.10$ is that of the unobscured AGN.
The values of $k_{W1}$ and $k_{W2}$ were derived assuming the standard extinction curve of the Galactic diffuse ISM \citep{Fitzpatrick99} and $R_V=3.1$.

The typical uncertainty of $A_{V,\mathrm{M22}}$ is $\sigma_{A_{V,\mathrm{M22}}}=7.7$ mag in this study, which is slightly better than that in \cite{Mizukoshi22} owing to the increase in the epoch data of \textit{WISE}.
However, this uncertainty is still large particularly for AGNs with small dust extinction.
To estimate $A_V$ of unobscured AGNs more precisely, we calculated $A_V$ following the method of \cite{Shimizu18} for AGNs with $A_{V,\mathrm{M22}}<15$ mag when broad H$\alpha$ flux data was available:

\begin{equation}
    A_{V,\mathrm{S18}}=\frac{1}{0.83}\times \left[-\frac{5}{2}\log \left(\frac{L_{\mathrm{bH\alpha,obs}}}{L_{\mathrm{bH\alpha,intr}}}\right)\right].
\end{equation}
Here, 0.83 was derived based on the empirical extinction law from \cite{Wild11}:

\begin{equation}
    \frac{A_{\lambda}}{A_V}=0.6(\lambda/5500)^{-1.3}+0.4(\lambda/5500)^{-0/7},
\end{equation}
and $\lambda(\mathrm{H\alpha})=6563${\AA}.
The observed broad H$\alpha$ line luminosity $L_{\mathrm{bH\alpha,obs}}$ can be calculated from its observed flux $f_{\mathrm{bH\alpha}}$ as $L_{\mathrm{bH\alpha,obs}}=f_{\mathrm{bH\alpha}}\times4\pi D^2$, where $D$ is the luminosity distance of each target.
The intrinsic broad H$\alpha$ line luminosity $L_{\mathrm{bH\alpha,intr}}$ was estimated from $L_{\mathrm{14-150keV,intr}}$ using the empirical relation from \cite{Shimizu18}:
\begin{equation}
    \log L_{\mathrm{bH\alpha,intr}}=1.06\log L_{\mathrm{14-150keV,intr}}-4.32.
\end{equation}

\begin{figure}
    \includegraphics[width=\linewidth]{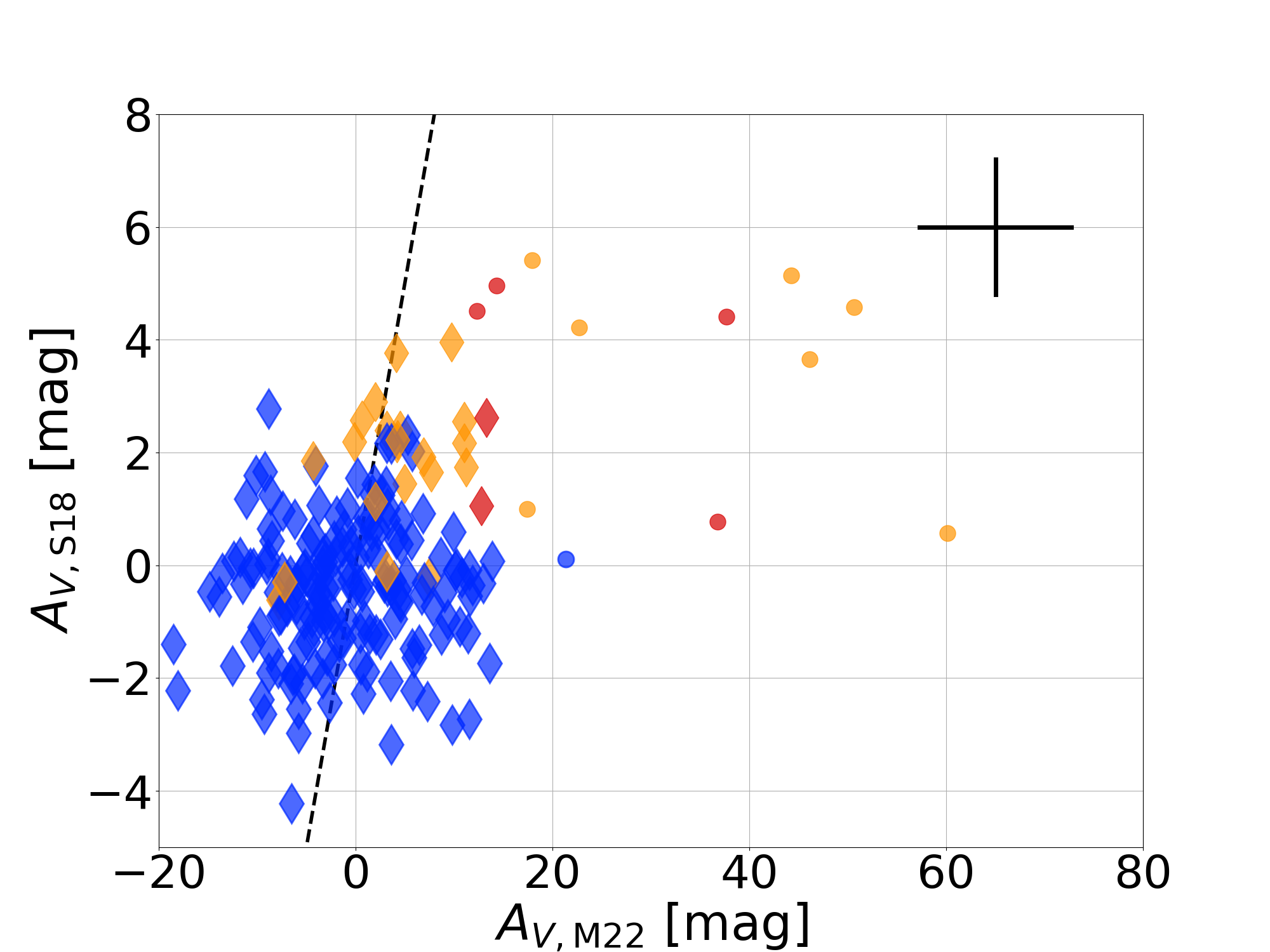}\par 
\caption{Comparison between the dust extinction based on the method of Mizukoshi et al. (2022) ($A_{V,\mathrm{M22}}$) and that based on Shimizu et al. (2018) ($A_{V,\mathrm{S18}}$).
The colours indicate the Seyfert type of each target.
The circles indicate the target to which we adopted $A_{V,\mathrm{M22}}$ and the diamonds indicate the target to which we adopted $A_{V,\mathrm{S18}}$. 
The black dashed line indicates the 1:1 relation.
The cross in the upper right represents the $\pm1\sigma$ error.}
\label{fig:Av-Av}
\end{figure}

The uncertainty of $A_{V,\mathrm{S18}}$ is $\sigma_{A_{V,\mathrm{S18}}}=1.2$ mag, which is much smaller than $\sigma_{A_{V,\mathrm{M22}}}$; however, $A_{V,\mathrm{S18}}$ possibly saturates at  approximately $5$ mag \citep{Xu20,Mizukoshi22}.
This trend is observed in Fig. \ref{fig:Av-Av}, which shows the comparison between $A_{V,\mathrm{M22}}$ and $A_{V,\mathrm{S18}}$ of our sample.
Although $A_{V,\mathrm{M22}}$ and $A_{V,\mathrm{S18}}$ are consistent with each other within the error for most unobscured AGNs, almost all obscured AGNs with large $A_{V,\mathrm{M22}}$ exhibit $A_{V,\mathrm{S18}}\sim 5$ mag regardless of $A_{V,\mathrm{M22}}$.
We therefore adopted $A_{V,\mathrm{S18}}$ for targets that satisfy both $A_{V,\mathrm{M22}}<15$ mag and $A_{V,\mathrm{S18}}<4$ mag.
There were six Sy2 AGNs for which $A_{V,\mathrm{S18}}$ can be measured. 
All of these targets have the data of $L_{\mathrm{bH\alpha,obs}}$ in BASS DR1. 
Half of them are actually classified as Sy1.9s in BASS DR1, while the remaining half are classified as Sy2s in it.

\begin{figure}
    \includegraphics[width=\linewidth]{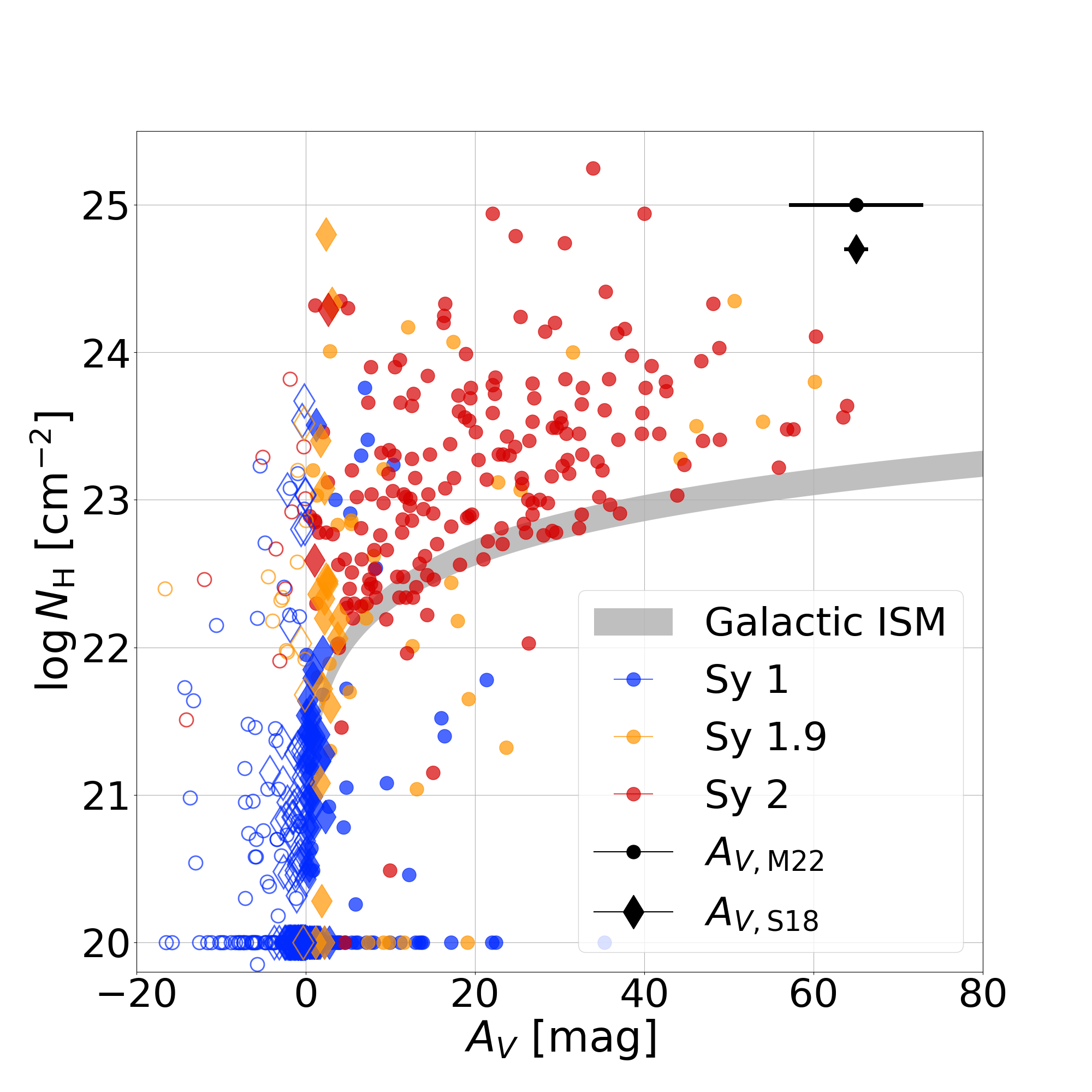}\par 
\caption{Comparison between the dust extinction $A_V$ and the gas column density $N_{\mathrm{H}}$.
The colours indicate the Seyfert type of each target and open markers with coloured edges indicate targets with $A_{V,\mathrm{M22/S18}}<0$ mag.
Circles indicate targets to which the $A_{V,\mathrm{M22}}$ is first adopted and diamonds indicate targets to which the $A_{V,\mathrm{S18}}$ is first adopted.
The gray band indicates the typical relation between $A_V$ and $N_{\mathrm{H}}$ of the Galactic diffuse ISM \citep{Predehl95,Nowak12}.
The black sediments in the upper right indicate the typical uncertainty of $A_{V,\mathrm{M22}}$ (upper side) and that of $A_{V,\mathrm{S18}}$ (lower side).}
 \label{fig:Av-NH}
\end{figure}

Figure \ref{fig:Av-NH} presents a comparison between $A_V$ and $N_{\mathrm{H}}$ of the targets.
The behaviour of our samples on this $A_V$--$N_{\mathrm{H}}$ diagram is nearly identical to that presented in \cite{Mizukoshi22} and other previous studies \citep[e.g.][]{Burtscher16,Shimizu18}.
In Fig.\ref{fig:Av-NH}, there is a non-trivial amount of Sy1.9, Sy2 AGNs with $A_{V,\mathrm{M22}}\lesssim0$ mag and this may be owing to a relatively large uncertainty of $A_{V,\mathrm{M22}}$.  
Most of these apparent unobscured Sy1.9, Sy2 AGNs are actually within the $2\sigma_{A_{V,\mathrm{M22}}}$ from $A_{V,\mathrm{M22}}\sim5$ mag.
In $A_V=5$ mag, the optical emission is attenuated by approximately two orders of magnitude, hence the detection of broad optical emission lines of these targets is challenging. 
Consequently, many AGNs with $A_V=5$ mag are thought to be classified as Sy1.9 or Sy2.
There are only three Sy1.9, Sy2 AGNs with $A_{V,\mathrm{M22}}$ smaller than $5$ mag by more than $2\sigma_{A_{V,\mathrm{M22}}}$.

There are many Sy1 AGNs and certain Sy1.9, Sy2 AGNs which show $A_{V,\mathrm{M22}}<0$ mag or $A_{V,\mathrm{S18}}<0$ mag owing to their uncertainty.
We adopted a lower limit of $A_V=0$ mag for these targets simply because $A_V$ should be physically larger than zero.
Furthermore, we assumed $A_V$ of targets which were distributed below the Galactic ISM relation in Fig. \ref{fig:Av-NH}, or targets with a smaller $N_{\mathrm{H}}/A_V$ than that of the Galactic ISM, that is, $[N_{\mathrm{H}}/A_V]_{\mathrm{Gal.}}$, to be $A_V= A_{V,\mathrm{Gal.}}\equiv N_{\mathrm{H}}/[N_{\mathrm{H}}/A_V]_{\mathrm{Gal.}}$ because the dusty and dust-free gas components cannot be separated for these targets in this study (Sec. \ref{subsec:dust-free NH separation}).
Table \ref{tab:sample number} summarizes the number of targets for which $A_{V,\mathrm{M22}}$, $A_{V,\mathrm{S18}}$, $A_V=0$ mag, or $A_V=A_{V,\mathrm{Gal.}}$ were adopted for each Seyfert type.

\defcitealias{smith2014}{Paper~I}
\begin{table}
\centering
 \caption{Number of targets for each Seyfert type and adopted dust extinction.}
 \label{tab:sample number}
 \begin{tabular}{lccccc}
  \hline
  Sy type & $A_{V,\mathrm{M22}}$ & $A_{V,\mathrm{S18}}$ & $A_V=0$ mag & $A_{V,\mathrm{Gal.}}$ & total\\

  \hline
Sy1 & 11 & 20 & 200& 87&318\\
Sy1.9 & 21 & 12 & 17& 18&68\\
Sy2 & 167 & 2 & 11 & 23&203\\

  \hline
total & 199 & 34 & 228 & 128& 589 \\

  \hline
\end{tabular}
\end{table}

\subsection{Separation of dusty and dust-free gas components}
\label{subsec:dust-free NH separation}

\begin{figure*}
    \includegraphics[width=\linewidth]{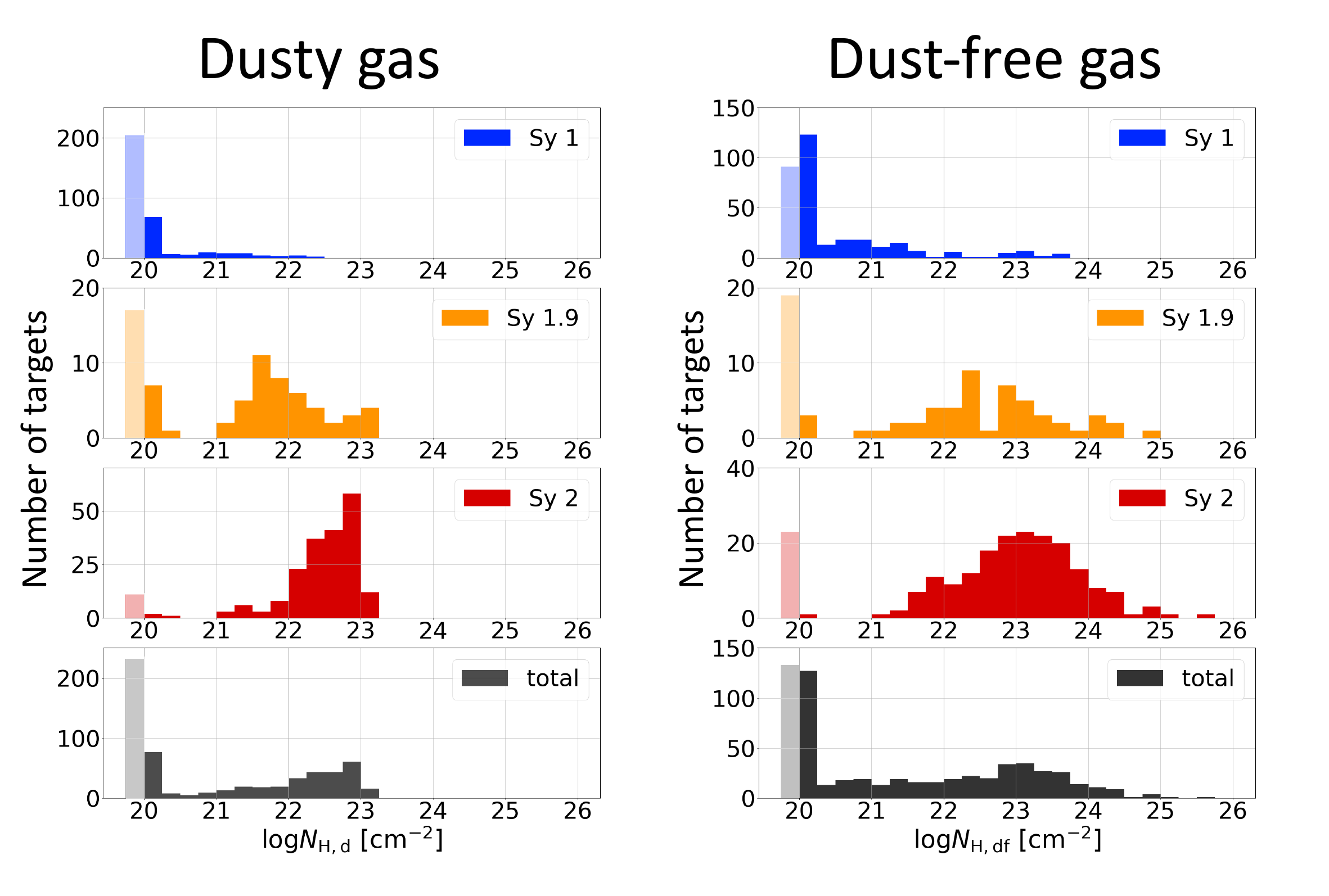}\par 
\caption{Histograms of the column density of the dusty gas $N_{\mathrm{H,d}}$ (left panel) and the dust-free gas $N_{\mathrm{H,df}}$ (right panel) for each Seyfert type and the total sample. 
The colours indicate the Seyfert types in the same way as in Fig. \ref{fig:Av-NH}.
In each dusty and dust-free gas histogram, we show the bar which indicates the target with $A_V=0$ mag and $\log N_{\mathrm{H,df}}\ [\mathrm{cm^{-2}}]<20$, respectively, in this study with pale colours.
We note that the column density in the range of $\log N_{\mathrm{H,d/df}}\ [\mathrm{cm^{-2}}]<22$ should be thought as the upper limit for a conservative discussion (see Sec.\ref{subsec:dust-free NH separation}).}
 \label{fig:NH-hist}
\end{figure*}

We separated $N_{\mathrm{H}}$ attributed to the dusty gas ($N_{\mathrm{H,d}}$) and dust-free gas ($N_{\mathrm{H,df}}$) assuming that the dusty gas surrounding the AGN typically have the same $N_{\mathrm{H}}/A_V$ ratio as that of the Galactic ISM.
Some previous studies found that the extinction curve of the AGN is rather flat compared to the Galactic one \citep[e.g.][]{Maiolino01,Gaskell04} and the $N_{\mathrm{H}}/A_V$ ratio of the AGN may be larger than that of the Galactic ISM.
These studies indicated that this is because the dust grain size is larger in the vicinity of the AGN compared to the Galactic ISM \citep[e.g.][]{Imanishi01,Maiolino01b}.
On the other hand, \cite{Baskin18} suggested that the change in the dust opacity owing to the change in the dust grain size is at most only a factor of two in the NIR band.
Therefore, the different assumptions of the $N_{\mathrm{H}}/A_V$ ratio are considered to have only a minor effect on the result of this study.

In this study, we calculated $N_{\mathrm{H,d}}$ as below:

\begin{equation}
    N_{\mathrm{H,d}}=A_V\times[N_{\mathrm{H}}/A_V]_{\mathrm{Gal.}},
\label{eq:NHdy}
\end{equation}
where $[N_{\mathrm{H}}/A_V]_{\mathrm{Gal.}}=(1.79-2.69)\times10^{21}\ \mathrm{[cm^{-2}\ mag^{-1}]}$ \citep{Predehl95,Nowak12}.
We here used $[N_{\mathrm{H}}/A_V]_{\mathrm{Gal.}}=(2.24\pm0.45)\times10^{21}\ \mathrm{[cm^{-2}\ mag^{-1}]}$ in our calculations for simplicity.
We then derived $N_{\mathrm{H,df}}$ by subtracting $N_{\mathrm{H,d}}$ from the observed $N_{\mathrm{H}}$ in BASS DR1 catalogue:

\begin{equation}
    N_{\mathrm{H,df}}= N_{\mathrm{H}} - N_{\mathrm{H,d}}.
\label{eq:NHdf}
\end{equation}

For the targets with $A_V=0$ mag, we assumed $N_{\mathrm{H,df}}$ to be equal to the observed $N_{\mathrm{H}}$ based on Eq. (\ref{eq:NHdf}).
We cannot calculate $N_{\mathrm{H,df}}$ with Eq. (\ref{eq:NHdf}) for the target with a smaller $N_{\mathrm{H}}/A_V$ than $[N_{\mathrm{H}}/A_V]_{\mathrm{Gal.}}$, or the targets with $A_V=A_{V,\mathrm{Gal.}}$; thus, we assumed $N_{\mathrm{H,df}}$ to be $\log N_{\mathrm{H,df}}\ [\mathrm{cm^{-2}}]=20$ for these targets as an upper limit.
We also adopted $\log N_{\mathrm{H,df}}\ [\mathrm{cm^{-2}}]=20$ for targets whose calculated $N_{\mathrm{H,df}}$ is smaller than this value.

The uncertainty of $N_{\mathrm{H,df}}$ is attributed to those of $N_{\mathrm{H}}$, $A_V$, and $[N_{\mathrm{H}}/A_V]_{\mathrm{Gal.}}$.
According to \cite{Ricci17c}, the $1\sigma$ error of $\log N_{\mathrm{H}}$ is calculated as 0.11 dex for targets with $\log N_{\mathrm{H}}\ [\mathrm{cm^{-2}}]>24$ and  0.04 dex for targets with $20<\log N_{\mathrm{H}}\ [\mathrm{cm^{-2}}]<24$.
Considering a simple error propagation, the $1\sigma$ error of $\log N_{\mathrm{H,df}}$ can be calculated as $0.11$ dex for targets with $\log N_{\mathrm{H}}\ [\mathrm{cm^{-2}}]>24$ using $\log N_{\mathrm{H}}[\mathrm{cm^{-2}}]=24.3$ and $A_V=25$ mag, which are the median of $N_{\mathrm{H}}$ and $A_V$ of these targets.
Similarly, the $1\sigma$ error of $\log N_{\mathrm{H,df}}$ can be calculated as $0.24$ dex for targets with $22\leq\log N_{\mathrm{H}}\ [\mathrm{cm^{-2}}]<24$ using $\log N_{\mathrm{H}}[\mathrm{cm^{-2}}]=23.0$ and $A_V=12.2$ mag.
For targets with $\log N_{\mathrm{H}}\ [\mathrm{cm^{-2}}]<22$, the obscuration may be primarily owing to the ISM in their host galaxies \citep{Fabian08,Fabian09,Ricci17c}; hence, $N_{\mathrm{H,df}}$ of these targets are considered to be the upper limit in a conservative discussion.

We summarized the main properties of the targets taken from both BASS DR1 \citep{Koss17,Ricci17b} and BASS DR2 \citep{Koss22a,Koss22b}, and results of our calculation in Tab. \ref{tab:result}.

\begin{table*}
 \caption{Properties of our final samples.}
 \label{tab:result}
 \begin{tabular}{lllllllllllll}
  \hline
  (1)&(2)&(3)$^a$&(4)$^a$&(5)$^a$&(6)$^b$&(7)$^c$&(8)$^{d}$&(9)&(10)&(11)&(12)\\
   BAT ID& Counterpart & Redshift & Sy type &  $\log f_{\mathrm{Edd}}$ & $\log N_{\rm{H}}$& flag & $A_V$ & $A_{V,\mathrm{M22}}$ & $\sigma_{A_{V,\mathrm{M22}}}$ & $A_{V,\mathrm{S18}}$ & $\log N_{\mathrm{H,df}}$ \\
  &&&&&(cm$^{-2}$)&&(mag)&(mag)&(mag)&(mag)&(cm$^{-2}$)\\
  \hline
1&2MASXJ00004876-0709117&0.0375&Sy1.9&-1.33&22.19& 2&4.0&9.7&7.7&4.0&21.82\\
2&2MASXJ00014596-7657144&0.0585&Sy1&-0.82&20.0& 3&0.0&-10.2&7.1&---&20\\
3&NGC7811&0.0255&Sy1&-0.86&20.0& 4&0.0&1.6&7.4&---&20\\
4&2MASXJ00032742+2739173&0.0398&Sy2&-1.55&22.86& 1&1.0&1.0&6.4&---&22.85\\
6&Mrk335&0.0259&Sy1&-1.17&20.48& 3&0.0&-9.3&7.0&-2.6&20.48\\
7&SDSSJ000911.57-003654.7&0.0733&Sy2&-1.67&23.56& 1&18.8&18.8&8.3&---&23.51\\
10&LEDA1348&0.0958&Sy1.9&-1.7&21.98& 3&0.0&-2.3&6.4&---&21.98\\
14&LEDA433346&0.0632&Sy1&-0.97&20.0& 3&0.0&-1.3&6.6&---&20\\
16&PG0026+129&0.142&Sy1&-0.98&20.0& 3&0.0&-6.3&7.2&---&20\\
19&RHS3&0.0743&Sy1&-1.41&20.0& 3&0.0&-9.7&8.5&---&20\\

  \hline
\multicolumn{10}{l}{\footnotesize $^a$ Data are taken from BASS DR2 catalogue \citep{Koss22a,Koss22b}.}\\
\multicolumn{10}{l}{\footnotesize $^b$ Data are taken from BASS DR1 catalogue \citep{Koss17,Ricci17b}.}\\
\multicolumn{10}{l}{\footnotesize $^c$ The flag to identify the dust extinction we adopted in this study. 1 is $A_{V,{\mathrm{M22}}}$, 2 is $A_{V,{\mathrm{S18}}}$, 3 is $A_{V}=0$ mag, and 4 is $A_{V,{\mathrm{Gal.}}}$.}\\
\multicolumn{10}{l}{\footnotesize $^d$ The dust extinction we adopted in this study.}\\
\multicolumn{10}{l}{\footnotesize (This table is available in its entirety in machine-readable form.)}\\
\end{tabular}
\end{table*}


\section{results}
\label{sec:results}

\subsection{Distribution of the column density of the dusty and dust-free gas}
\label{subsec:NH distribution}

The left panel of Fig. \ref{fig:NH-hist} shows the histograms of $N_{\mathrm{H,d}}$ for each Seyfert type and the total sample of this study.
In this figure, we gather all samples with $A_{V,\mathrm{M22}}$,  $A_{V,\mathrm{S18}}$, or $A_{V,\mathrm{Gal.}}$, and the targets with $A_V=0$ mag are shown using pale colours.
Most $N_{\mathrm{H,d}}$ of the Sy1 AGN exhibits $\log N_{\mathrm{H,d}}\ [\mathrm{cm^{-2}}]\lesssim22$, or $A_V\lesssim5$ mag if we adopt $[N_{\mathrm{H}}/A_V]_{\mathrm{Gal.}}$, which may be owing to the ISM in the host galaxy.
For Sy2 AGNs, the histogram exhibits a peak at $\log N_{\mathrm{H,d}}\ [\mathrm{cm^{-2}}]\sim23$ and there are relatively few targets with a larger $N_{\mathrm{H,d}}$.

The right panel of Fig. \ref{fig:NH-hist} shows the histograms of $N_{\mathrm{H,df}}$ for each Seyfert type and the total sample.
Similarly to the dusty gas, most Sy1 AGNs show $\log N_{\mathrm{H,df}}\ [\mathrm{cm^{-2}}]\lesssim22$.
Although the histogram of Sy2 AGNs exhibits a peak at $\log N_{\mathrm{H,df}}\ [\mathrm{cm^{-2}}]\sim23$, which is also similar to that of $N_{\mathrm{H,d}}$, there are many Sy2 AGNs with a large $N_{\mathrm{H,df}}$ of $\log N_{\mathrm{H,df}}\ [\mathrm{cm^{-2}}]>23$ or even the Compton-thick absorption of $\log N_{\mathrm{H,df}}\ [\mathrm{cm^{-2}}]\gtrsim24$, which is a different behaviour to that of the $N_{\mathrm{H,d}}$ histogram.

The typical column density of both the dusty and dust-free gas is different in each Seyfert type.
Based on the unified model, this difference may be owing to the orientation effect of these gas structures.
Furthermore, the histograms of $N_{\mathrm{H,d}}$ and $N_{\mathrm{H,df}}$ of the total sample exhibit different distributions from each other, wherein there are few targets with $\log N_{\mathrm{H,d}}\ [\mathrm{cm^{-2}}]>23$ ($A_V\gtrsim65$ mag), although there are a certain number of targets with $\log N_{\mathrm{H,df}}\ [\mathrm{cm^{-2}}]>23$.
To confirm the difference between these distributions statistically, we performed the Kolmogorov–Smirnov test.
The resulting $p$-value was $7\times10^{-41}$, which suggests that these distributions are very likely to be different.
In this study, we excluded 62 targets from our final sample because of a weak correlation between $W1$ and $W2$ fluxes or low accuracy of the FVG.
Most of these targets were Sy1.9 or Sy2 AGNs and the mean total gas column density was as large as $\log N_{\mathrm{H}}\ [\mathrm{cm^{-2}}]\sim23.5$. Thus the dust extinction was expected to be large as well.
This possible heavy extinction may decrease the observed flux time variation and worsen the correlation between $W1$ and $W2$ fluxes.
Although these targets with possible heavy dust obscuration are expected to contribute to the range of $\log N_{\mathrm{H,d}}\ [\mathrm{cm^{-2}}]>23$ or even the Compton-thick dust obscuration ($\log N_{\mathrm{H,d}}\ [\mathrm{cm^{-2}}]>24$) in the $N_{\mathrm{H,d}}$ histogram of the Sy1.9 or  Sy2 AGN, these targets comprise approximately 10\% of our final sample. Thus, these targets exert only a minor effect on the distribution of the $N_{\mathrm{H,d}}$ histogram and the final result of this study (Sec.\ref{subsubsec:Comparison with Ricci+}).

\cite{Silver22} performed a detailed X-ray spectral modeling for four Compton-thick AGN candidates; 2MASX J02051994-0233055 (hereafter 2MASX J0205), IC2227, 2MASX J04075215-6116126 (hereafter 2MASX J0407), and ESO362-8 using the data simultaneously obtained by XMM-Newton \citep{Jansen01} and Nuclear Spectroscopic Telescope Array \citep[NuSTAR,][]{Harrison13}.
They used several physically-motivated torus models to investigate the accurate characteristics of the obscuration of these targets, and concluded that 2MASX J0205 was an unobscured AGN ($\log N_{\mathrm{H}}\ [\mathrm{cm^{-2}}]<22$), both 2MASX J0407 ($\log N_{\mathrm{H}}\ [\mathrm{cm^{-2}}]\sim23.5$) and IC2227 ($\log N_{\mathrm{H}}\ [\mathrm{cm^{-2}}]\sim23.8$) were Compton-thin AGNs, and ESO362-8 was a bona fide Compton-thick AGN ($\log N_{\mathrm{H}}\ [\mathrm{cm^{-2}}]>24$).

We performed the same analysis in this study for these four targets of \cite{Silver22} to assess the validity of our analyses.
For 2MASX J0205, the WISE light curve exhibited clear time variation and we derived $A_{V,\mathrm{M22}}=-2.5\pm7$ mag, which was consistent with \cite{Silver22}.
IC2227 similarly shows the WISE flux time variation and we derived $A_{V,\mathrm{M22}}=-2.1\pm8$ mag.
For 2MASX J0407, we derived $A_{V,\mathrm{M22}}=31\pm7$ mag, which is equivalent to $\log N_{\mathrm{H,d}}\ [\mathrm{cm^{-2}}]=22.8$.
The dust extinction of both IC2227 and 2MASX J0407 was considerably smaller than that estimated from $N_{\mathrm{H}}$ in \cite{Silver22}.
This can be explained that the physically-motivated torus model generally considers only the dusty torus as an obscuring structure; hence, even if there is a large contribution of dust-free gas to $N_{\mathrm{H}}$, it does not distinguish the obscuration by dusty torus and dust-free gas. 
Consequently, the discrepancy between the results of \cite{Silver22} and our analyses for these two targets does not contradict to the dust-free gas scenario.
Furthermore, \cite{Silver22} noted the possibility of the $N_{\mathrm{H}}$ time variation in IC2227. This was also consistent with the dust-free gas scenario wherein the $N_{\mathrm{H}}$ time variation was attributed to the eclipse of the dust-free gas cloud \citep[e.g.,][]{Burtscher16}. 
Finally, the WISE light curve of ESO362-8 did not exhibit clear time variation and we cannot measure its $A_{V,\mathrm{M22}}$.
This result suggests that ESO362-8 has heavy dust extinction of at least $A_{V}>65$ mag, which is consistent with the result of \cite{Silver22}. 

A possible caveat for the small number of the target with $\log N_{\mathrm{H,d}}\ [\mathrm{cm^{-2}}]\gtrsim23$ is that we may not be able to observe the NIR radiation in the heavily-obscured line of sight with $\log N_{\mathrm{H,d}}\ [\mathrm{cm^{-2}}]>23$ and the dust IR radiation from out of the line of sight leaks to be observed \citep[e.g.][]{Schartmann14,Matsumoto22}, resulting in the underestimation of $A_V$.
This is partly because the observed NIR radiation is averaged from that originating from the hot dust region, which is spatially more extended than the central X-ray source.
Furthermore, if the hot dust is primarily distributed near the surface of the anisotropic dust-sublimation region \citep[e.g.][]{Kawaguchi10}, the method of \cite{Mizukoshi22} cannot measure $A_V$ in the truly edge-on direction, which may also result in the underestimation of $A_V$.

Thus, we conclude that (i) both the dusty and dust-free gas structures  exhibit a difference in the column density distribution for each Sy type, which may be owing to the orientation effect,
and (ii) the column-density distribution of the dusty and dust-free gas may differ in the range of $\log N_{\mathrm{H,d/df}}\ [\mathrm{cm^{-2}}]\gtrsim23$, which is primarily observed for the Sy1.9 or Sy2 AGN.

\subsection{Properties of the dusty gas structure}
\label{subsec:dusty gas}

\subsubsection{Effects of the radiation pressure on the dusty gas}
\label{subsubsec:dusty gas vs fEdd}

\begin{figure*}
    \includegraphics[width=0.7\linewidth]{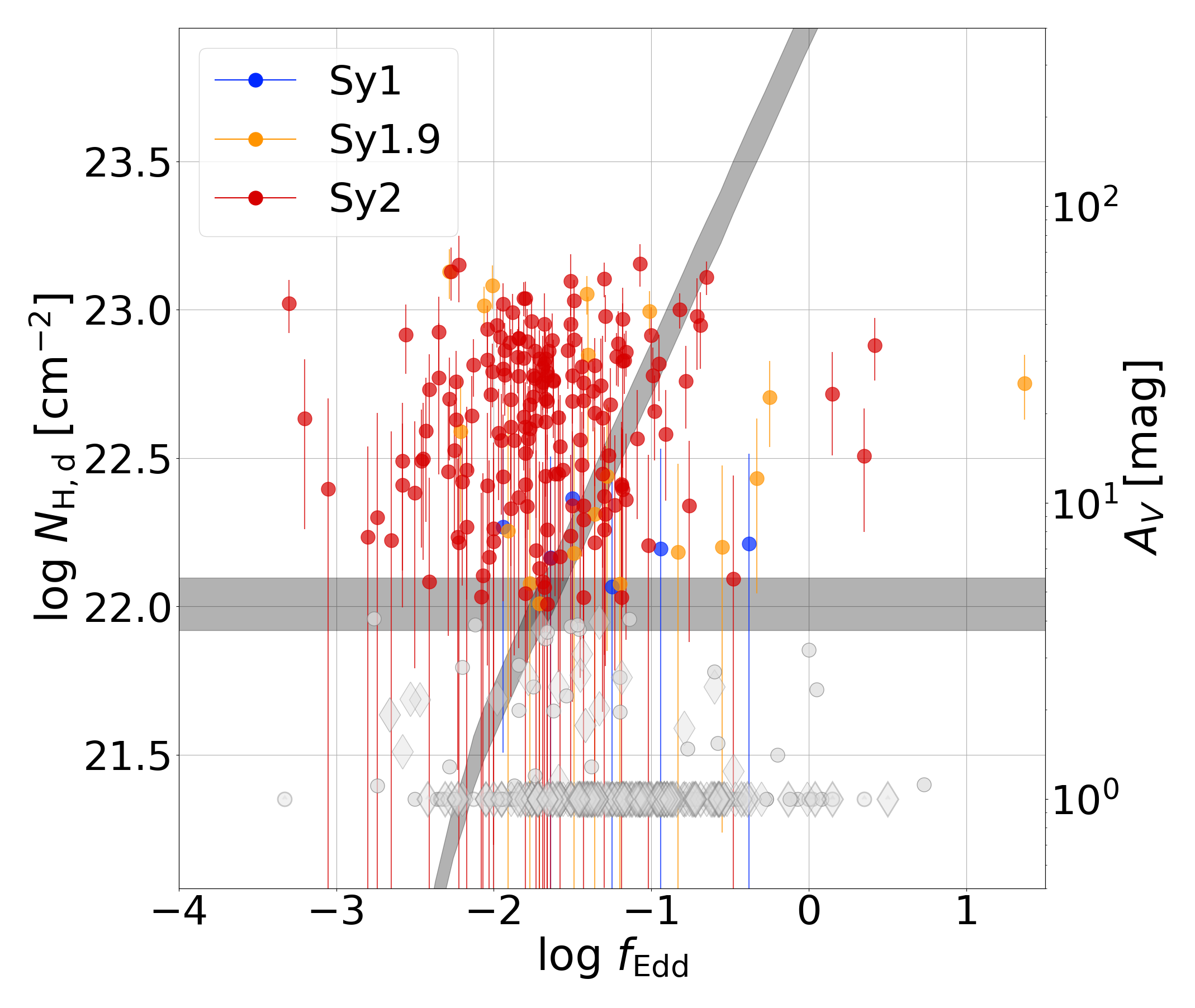}\par 
\caption{Comparison between Eddington ratio $f_{\mathrm{Edd}}$ and dust extinction. The symbols and colours follow those in Fig. \ref{fig:Av-NH}, while we show the targets with $\log N_{\mathrm{H,d}}\ [\mathrm{cm^{-2}}]<22$ as gray markers because their $N_{\mathrm{H,d}}$ are thought to be the upper limits (see Sec. \ref{subsec:dust-free NH separation} for detail). 
The diagonal gray band indicates the relation of $L_{\mathrm{bol}}=L_{\mathrm{Edd}}^{\mathrm{eff}}$ \citep[taken from][]{Fabian08}, and the horizontal gray band is comparable to $\log N_{\mathrm{H,d}}\ \mathrm{[cm^{-2}]}=22$.
The width of these gray bands indicates the uncertainty which is attributed to that of the $[N_{\mathrm{H}}/A_V]_{\mathrm{Gal.}}$.
In this figure, we show all targets with $A_V<1$ mag at $A_V=1$ mag as the upper limit because these small $A_V$ are comparable to $A_V=0$ mag within the uncertainty. }
\label{fig:Av_Eratio}
\end{figure*}

Figure \ref{fig:Av_Eratio} shows a comparison between $A_V$ and $f_{\mathrm{Edd}}$ of our sample.
We converted both the curve that represents $f_{\mathrm{Edd}}^{\mathrm{eff}}$ for different $N_{\mathrm{H}}$ \citep{Fabian08} and the horizontal line of $\log N_{\mathrm{H}}\ [\mathrm{cm^{-2}}]=22$, which shows the lower boundary of the forbidden region in \cite{Ricci17c}, using the $[N_{\mathrm{H}}/A_V]_{\mathrm{Gal.}}$.
Both are shown as the grey bands in Fig. \ref{fig:Av_Eratio} and their width represents the uncertainty which originates from that of the $[N_{\mathrm{H}}/A_V]_{\mathrm{Gal.}}$.
\cite{Ishibashi18} calculated $f_{\mathrm{Edd}}^{\mathrm{eff}}$ considering the IR radiation trapping and suggested that, while $f_{\mathrm{Edd}}^{\mathrm{eff}}$ behaved similarly to that of the result of \cite{Fabian08} in $\log N_{\mathrm{H}}\ [\mathrm{cm^{-2}}]\lesssim23$, IR-optically thick material ($\log N_{\mathrm{H}}\ [\mathrm{cm^{-2}}]\gtrsim23$) can be blown out even in sub-Eddington states.
In this study, most samples are distributed in the range of $\log N_{\mathrm{H,d}}\ [\mathrm{cm^{-2}}]\lesssim23$, where the effect of the IR radiation trapping is thought to be small.
In addition, although the launching point of $f_{\mathrm{Edd}}^{\mathrm{eff}}$ in the $N_{\mathrm{H}}$-$f_{\mathrm{Edd}}$ diagram is different between \cite{Ishibashi18} and \cite{Fabian08}, this is primarily owing to the difference in the ratio of the effective cross section of the dusty gas to the Thomson cross section they assumed or derived; thus, it is not a fundamental difference of their results.
Therefore, we did not explicitly perform comparisons with $f_{\mathrm{Edd}}^{\mathrm{eff}}$ presented in \cite{Ishibashi18}.

In Fig. \ref{fig:Av_Eratio}, there are few targets, approximately only 4\% of the total sample, in the forbidden region, and most Sy2 AGNs are distributed just out of or on the left boundary of the forbidden region.
While this is the same trend as in the $N_{\mathrm{H}}$--$f_{\mathrm{Edd}}$ diagram in previous studies \citep{Fabian08,Fabian09,Ricci17c,Ricci22}, the distribution of these obscured AGNs appears to be more concentrated in the $A_V$--$f_{\mathrm{Edd}}$ diagram and shows a clearer boundary of the forbidden region than in the $N_{\mathrm{H}}$--$f_{\mathrm{Edd}}$ diagram.
This is because the dust-free gas component was excluded and the vertical scatter of the distribution of these targets became smaller by using $A_V$ instead of $N_{\mathrm{H}}$.

This result supports the AGN evolutionary scenario that has been suggested in the literature based on the $N_{\mathrm{H}}$--$f_{\mathrm{Edd}}$ diagram \citep[e.g. ][]{Jun21,Ricci22}. (i) Both the $A_V$ and the $f_{\mathrm{Edd}}$ will increase in AGNs with active gas accretion and many obscured AGNs stay near the boundary of the forbidden region. (ii) Once they enter the forbidden region, the accreting material which obscures the central engine will be blown out by the radiation-driven dusty gas outflow. Finally, (iii) AGNs change from obscured ones to unobscured ones in a short time scale.

\subsubsection{dusty gas covering factor and its Eddington-ratio dependence}
\label{subsubsec:dusty gas covering factor}

\begin{figure*}
    \includegraphics[width=0.7\linewidth]{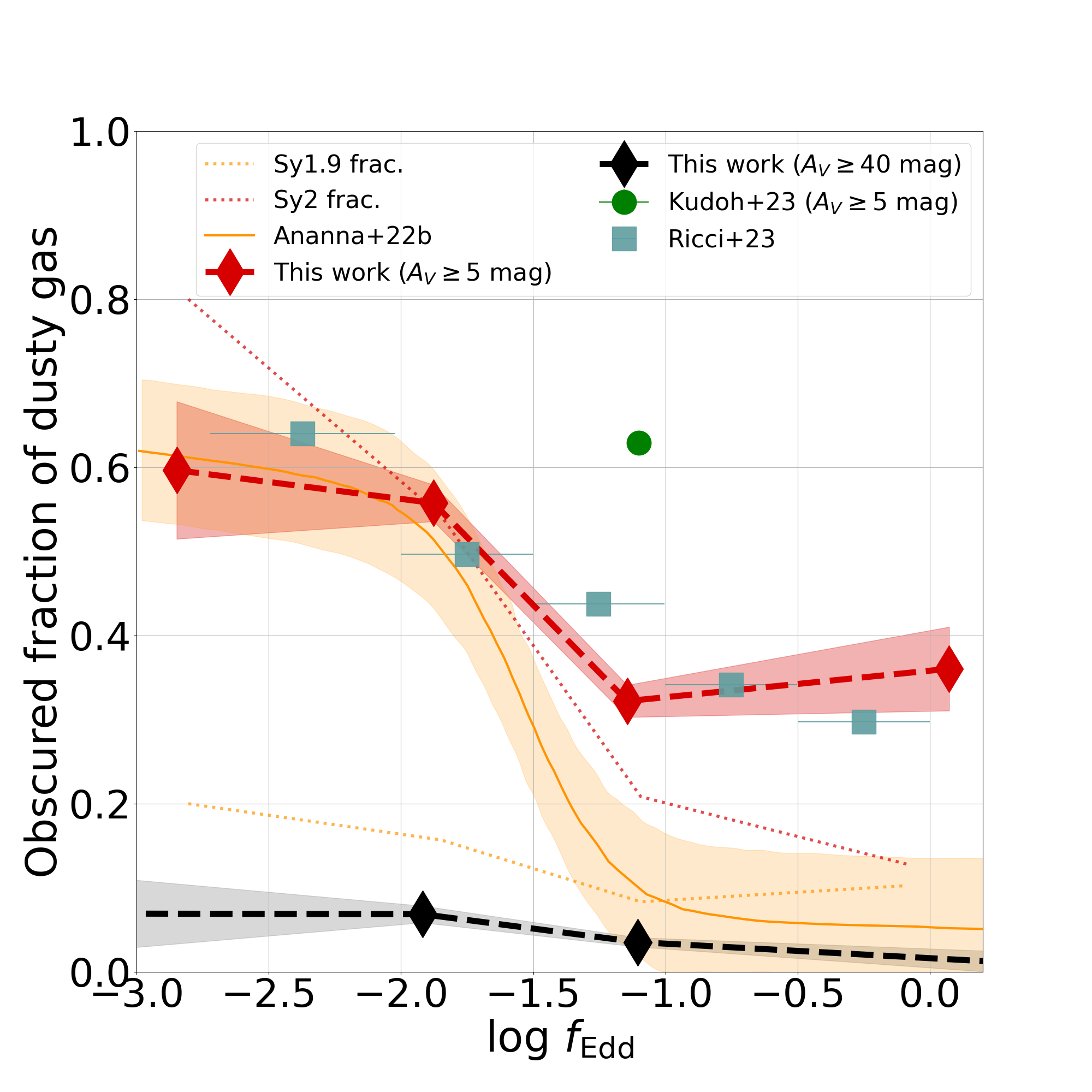}\par 
\caption{The fraction of the dust-obscured AGN as a function of the Eddington ratio.
The red and black diamonds connected with dashed lines indicate the fraction of the target with $A_V\geq5$ mag and that with $A_V\geq40$ mag, respectively, for each Eddington-ratio bin in this study, and the red and grey shaded area indicate their $1\sigma$ error, respectively. The red dotted line indicates the Sy2 AGN fraction of our final sample for each Eddington-ratio bin and the orange dotted line indicates that of the Sy1.9 AGN.  The green filled circle indicates the covering factor of the dusty gas with $A_V\geq5$ mag in the model of Kudoh et al. (2023). The dark blue squares indicate the dusty gas covering factor based on the ratio of the IR luminosity and the bolometric luminosity (Ricci et al. 2023). }
\label{fig:Av_fraction}
\end{figure*}

\begin{figure*}
    \includegraphics[width=0.7\linewidth]{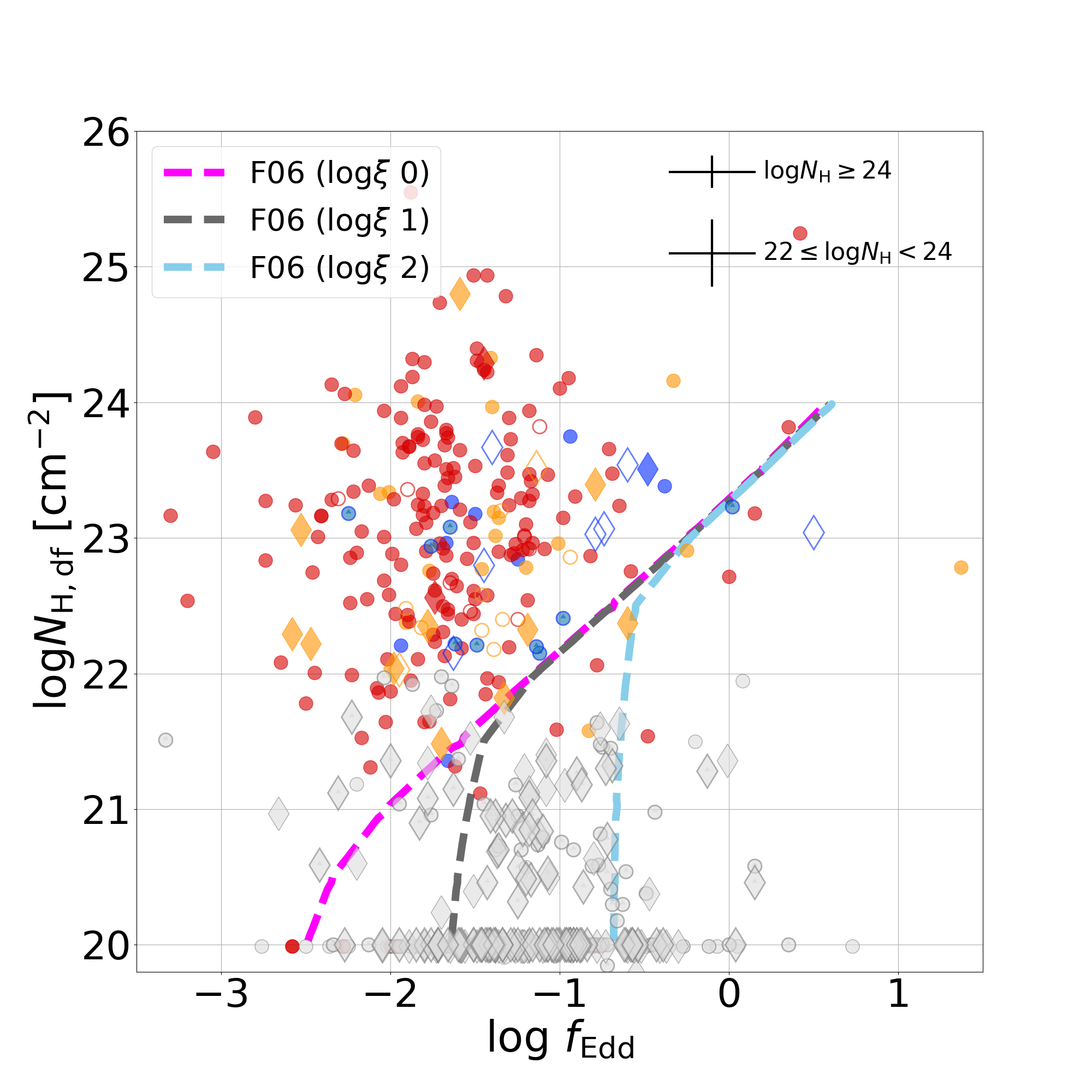}\par 
\caption{Comparison between Eddington ratio $f_{\mathrm{Edd}}$ and the column density of the dust-free gas $N_{\mathrm{H,df}}$. 
The symbols and colours follow those in Fig. \ref{fig:Av-NH}, while we here show the targets with $\log N_{\mathrm{H}}\ [\mathrm{cm^{-2}}]<22$ with gray markers. 
The error bars in the top right indicate $1\sigma$ error for targets in each $\log N_{\mathrm{H}}$ bin.
The magenta, gray, and cyan dashed lines indicate the relation of $L_{\mathrm{bol}}=L_{\mathrm{Edd}}^{\mathrm{eff}}$ for the dust-free gas with the ionization parameters of $\log \xi=0$, $1$, or $2$, respectively \citep{Fabian06}.}
\label{fig:dust-free-gas_Eratio}
\end{figure*}

We calculated the fraction of the AGN with either $A_V\geq5$ mag or $A_V\geq40$ mag in our sample.
Based on the AGN unified model, this obscured fraction is equivalent to the covering factor of the dusty gas that contributes to the same amount of dust extinction. 
The covering factor of the dusty gas with $A_V\geq5$ mag, which is equivalent to $\log N_{\mathrm{H,d}} [\mathrm{cm^{-2}}]\gtrsim22$ for the Galactic ISM, indicates an approximate total covering factor of the dusty torus, whereas that with $A_V\geq40$ mag, which is equivalent to $\log N_{\mathrm{H,d}} [\mathrm{cm^{-2}}]\gtrsim23$  for the Galactic ISM, exhibits the behaviours of the heavily-obscuring dusty gas component.
The method of \cite{Mizukoshi22} cannot measure a very large $A_V$ of $A_V\gtrsim65$ mag, or $\log N_{\mathrm{H,d}} [\mathrm{cm^{-2}}]\gtrsim23.2$ for the Galactic ISM; thus, we cannot directly investigate the Compton-thick dusty gas structure.
Nonetheless, as explained in Sec.\ref{subsec:NH distribution}, the covering factor of such Compton-thick dusty gas component can be very small (see also Sec.\ref{subsubsec:Comparison with Ricci+}).

In the calculation of the covering factor, we separated the sample into four bins of the Eddington ratio: (i) $\log f_{\mathrm{Edd}}<-2.5$ (15 samples), (ii) $-2.5\leq\log f_{\mathrm{Edd}}<-1.5$ (223 samples), (iii) $-1.5\leq\log f_{\mathrm{Edd}}<-0.5$ (312 samples), and (iv) $\log f_{\mathrm{Edd}}\geq-0.5$ (39 samples).
In this calculation, we first added to $A_{V,\mathrm{M22}}$ or $A_{V,\mathrm{S18}}$ of each target a random error following a normal probability distribution with $\sigma_{A_{V,\mathrm{M22}}}=7.7$ mag or $\sigma_{A_{V,\mathrm{S18}}}=1.2$ mag, respectively.
Thereafter, we calculated the fraction of targets with $A_V\geq5$ mag and those with $A_V\geq40$ mag.
To evaluate the uncertainty of this obscured fraction, we performed this calculation 1000 times and considered the average and the standard deviation of the total results of these calculations as the resulting covering factor and its $1\sigma$ error.

Figure \ref{fig:Av_fraction} shows the obscured fraction, or the covering factor of the dusty gas with $A_V\geq5$ mag (red dashed line) and those with $A_V\geq40$ mag (black dashed line).
The covering factor of the dusty gas with $A_V\geq5$ mag was $f_{\mathrm{C}}\sim0.6$ in $\log f_{\mathrm{Edd}}\lesssim-2$ and $f_{\mathrm{C}}\sim0.3$ in $\log f_{\mathrm{Edd}}\gtrsim-1$, and relatively constant in these $f_{\mathrm{Edd}}$ ranges within the error.
However, in the range of $-2<\log f_{\mathrm{Edd}}<-1$, the covering factor exhibited a clear drop.
This result suggests that dust-obscured AGNs enter the forbidden region and cause the dusty gas outflow typically in $-2<\log f_{\mathrm{Edd}}<-1$, resulting in the decrease of the covering factor of the dusty gas, while the dusty gas structure is relatively stable in lower or higher Eddington ratio. 
This is consistent with the outcome that the left boundary of the forbidden region, or $f_{\mathrm{Edd}}^{\mathrm{eff}}$, lies around $-2<\log f_{\mathrm{Edd,eff}}<-1$ in the range of $A_V\gtrsim5$ mag in Fig. \ref{fig:Av_Eratio}.
The covering factor of the dusty gas with $A_V\geq40$ mag was as small as $f_{\mathrm{C}}\sim0.05$ and nearly independent of the Eddington ratio within the error.
This result indicates that the dusty gas structure that contributes to the heavy dust extinction is not largely affected by the outflow.
This behaviour is similar to that of the Compton-thick gas structure with $\log N_{\mathrm{H}}\ [\mathrm{cm^{-2}}]>24$ \citep{Ricci17c}.
We summarized the calculated fraction of the AGN with $A_V\geq5$ mag and that with $A_V\geq40$ mag in Tab. \ref{tab:covering factor}.

In Fig. \ref{fig:Av_fraction}, we show the estimated intrinsic fraction of the Sy2 AGN as a function of the Eddington ratio \citep{Ananna22a,Ananna22b}.
They performed a Bayesian inference method to correct the effects of the Eddington bias \citep{Eddington1913} and the obscuration to the AGN.
For $\log f_{\mathrm{Edd}}<-2$, the intrinsic Sy2 AGN fraction was estimated to be $f_{\mathrm{C}}\sim0.6$ and nearly constant within the error.
The fraction then reduced to $f_{\mathrm{C}}\sim0.1$ at $\log f_{\mathrm{Edd}}\sim-1.5$ and then became constant again for $\log f_{\mathrm{Edd}}>-1$.
This behaviour of the intrinsic Sy2 AGN fraction shows qualitatively a good agreement with the obscured fraction with $A_V\geq5$ mag in this study within the error; whereas in $\log f_{\mathrm{Edd}}\gtrsim-2$, the intrinsic Sy2 AGN fraction shows a more steep drop and has a lower value in $\log f_{\mathrm{Edd}}\gtrsim-1$.
One possible reason for this difference is the Eddington bias. 
As the uncertainty of the Eddington ratio increases, the number of the AGN is generally overestimated in a higher $f_{\mathrm{Edd}}$ regime because the number of the AGN with a high Eddington ratio is intrinsically small.
Consequently, the obscured fraction of this study in the high-Eddington regime may exhibit certain overestimation.
Another possible explanation is that we should compare the combined fraction of Sy2 and Sy1.9 AGNs to the obscured fraction of this study, considering that approximately half of Sy1.9 AGNs in this study exhibited relatively large dust extinction of $A_V\gtrsim5$ mag, or $\log N_{\mathrm{H,d}}\ [\mathrm{cm^{-2}}]\gtrsim22$ (Fig.\ref{fig:NH-hist}).
We show the fraction of the Sy1.9 (orange dotted line) and Sy2 AGN (red dotted line) of our sample in each Eddington-ratio bin in Fig.\ref{fig:Av_fraction}.
The combined fraction of Sy1.9 and Sy2 AGNs in $\log f_{\mathrm{Edd}}\gtrsim-1.5$ is comparable to that of the AGN with $A_V\geq 5$ mag in this study, whereas the Sy2 AGN fraction of our final sample is actually comparable to that of \cite{Ananna22b} in the same Eddington-ratio regime.
However, this explanation cannot be applied to the low-Eddington regime because the Sy2 AGN fraction of this study (approximately $0.8$) for $\log f_{\mathrm{Edd}}\lesssim-2$ is larger than the obscured fraction in this study.
Moreover, the difference increases upon the addition of the Sy1.9 AGN fraction.
This large Sy2 AGN fraction compared to the obscured fraction may be owing to faint Sy1 AGNs with weak broad emission lines, which may be misclassified as Sy2 AGNs.



\cite{Kudoh23} performed a two-dimensional radiation-hydrodynamic simulation of the central sub-pc region of a modeled AGN with $\log f_{\mathrm{Edd}}\sim-1$, and derived gas column density profiles along the elevation angle for both the dusty  and dust-free gas.
In Fig. \ref{fig:Av_fraction}, we show the covering factor of the dusty gas with $\log N_{\mathrm{H,d}}\ [\mathrm{cm^{-2}}]\geq22$, or $A_V\geq5$ mag for the Galactic ISM, based on the column density profile of \cite{Kudoh23}.
The covering factor of \cite{Kudoh23} is much larger than our result at the similar Eddington ratio.
It is even larger than the total covering factor of Compton-thin and Compton-thick gas estimated with the \textit{Swift}/BAT catalogue \citep{Ricci17c} at the same Eddington ratio.
This discrepancy will be discussed in a forthcoming paper.

Recently, \cite{Ricci23} investigated the relation between the Eddington ratio and the dusty gas covering factor which was estimated based on the ratio of the IR luminosity to the bolometric luminosity using BASS AGN samples.
Consequently, they showed that the covering factor decreased in higher Eddington ratio from $f_{\mathrm{C}}\sim0.6$ at $\log f_{\mathrm{Edd}}\sim-2.5$ to $f_{\mathrm{C}}\sim0.3$ at $\log f_{\mathrm{Edd}}\sim0$ when they adopted an $f_{\mathrm{Edd}}$-dependent 2-10 keV bolometric correction \citep[][]{Vasudevan07}.
This result is consistent with those of this study within the uncertainty (dark blue squares in Fig.\ref{fig:Av_fraction}).
Although they suggested that the less steep decreasing trend of the IR-based covering factor compared to the previous X-ray study \citep{Ricci17c} may be owing to the effect of the IR emission from polar dust, this difference can be also explained by the fact that the relatively steep decreasing trend in the X-ray study may be primarily owing to the dust-free gas structure.
We present the covering factor of the dust-free gas and its Eddington-ratio dependence in Sec.\ref{subsubsec:dust-free gas covering factor}.


\subsection{Properties of the dust-free gas structure}
\label{subsec:dust-free gas}



\subsubsection{Effects of the radiation pressure on the dust-free gas}
\label{subsubsec:dust-free gas vs fEdd}

Figure \ref{fig:dust-free-gas_Eratio} presents a comparison of  $N_{\mathrm{H,df}}$ and $f_{\mathrm{Edd}}$ of our sample.
In this figure, most Sy1 AGNs are distributed in the range of $\log N_{\mathrm{H,df}}\ [\mathrm{cm^{-2}}]\lesssim22$, whereas Sy2 AGNs are primarily distributed in the range of  $\log N_{\mathrm{H,df}}\ [\mathrm{cm^{-2}}]\gtrsim22$.
As expected from the right panel of Fig. \ref{fig:NH-hist},  this result indicates that the typical column density of the dust-free gas in the line of sight varies depending on the Seyfert type.

The dust-free gas does not contain dust, hence the effective Eddington limit for the dusty gas cannot be applied when we consider the radiation-driven outflow of the dust-free gas.
\cite{Fabian06} used a radiative transfer code CLOUDY \citep{Ferland93} to calculate the effective cross section of the partially-ionized gas as a function of its $N_{\mathrm{H}}$.
Consequently, they found that partially-ionized dust-free gas also exhibited a much larger cross section than the Thomson cross section owing to photoelectric absorption, and it decreased in higher $N_{\mathrm{H}}$ in the same manner as that of the dusty gas. 
We show several lines at which the bolometric luminosity is equal to the effective Eddington limit of partially-ionized dust-free gas in Fig. \ref{fig:dust-free-gas_Eratio}.
Each line indicates a different ionization state which is parameterized by the ionization parameter $\xi\ [\mathrm{erg\ cm\ s^{-1}}]\equiv L/nr^2$, where $L$ is the ionizing luminosity, $n$ is the gas number density, and $r$ is the distance from the central source.
The curve representing higher $\xi$ shows a larger effective Eddington limit because the dust-free gas in a higher ionization state has a smaller cross section for radiation.
In Fig. \ref{fig:dust-free-gas_Eratio}, targets with a larger $f_{\mathrm{Edd}}$ than the effective Eddington limit of the dust-free gas with certain $\xi$ are expected to exhibit the radiation-driven outflow of the dust-free gas with this $\xi$, although this may not hold for the target with $\log N_{\mathrm{H,df}}\ [\mathrm{cm^{-2}}]\lesssim21$ because the gas with relatively small column density will be completely ionized with the radiation from the central source \citep[e.g.][]{Fabian08}.

Figure \ref{fig:dust-free-gas_Eratio} shows that almost all targets with $\log N_{\mathrm{H}}\ [\mathrm{cm^{-2}}]>22$ (coloured plots) are distributed in a lower-$f_{\mathrm{Edd}}$ regime compared to the effective Eddington limit of the dust-free gas with $\log \xi=0$.
This result indicates that the ionization parameter of the dust-free gas of these obscured AGNs is typically $\log \xi\sim0$.
In the range of $\log N_{\mathrm{H,df}}\ [\mathrm{cm^{-2}}]<22$, most targets exhibit $\log N_{\mathrm{H}}\ [\mathrm{cm^{-2}}]<22$; hence, $N_{\mathrm{H,df}}$ of these targets should be considered as the upper limit (Sec.\ref{subsec:dust-free NH separation}).
Furthermore, the column density in this study \citep{Ricci17b} was assumed to be owing to the cold neutral gas, hence it cannot be regarded as that of the (partially-) ionized gas.
This also renders these relatively small $N_{\mathrm{H,df}}$ even more uncertain.
Nevertheless, Fig.\ref{fig:dust-free-gas_Eratio} hints that most AGNs with $\log N_{\mathrm{H}}\ [\mathrm{cm^{-2}}]<22$ are distributed in the lower--$f_{\mathrm{Edd}}$ regime than the effective Eddington limit of the dust-free gas with $\log \xi=2$ and the upper limit of the ionization parameter of the dust-free gas is $\log \xi\sim2$ for unobscured AGNs.

Such estimation of $\xi$ enables us to constrain the physical scale of the dust-free gas structure.
\cite{Granato97} interpreted the observed large difference between the level of the dust extinction and the X-ray absorption as the effect of the dust-free gas inside the dust-sublimation zone.
\cite{Burtscher16} also suggested that the dust-free gas cloud was distributed in the BLR based on the short time scale of the observed $N_{\mathrm{H}}$ variation \citep[e.g.][]{Risaliti02,Bianchi09,Bianchi12}.
Furthermore, recent radiation-hydrodynamic simulation \citep{Kudoh23} showed that the physical scale of the dust-free gas region is similar or smaller ($\lesssim0.01$ pc) than the BLR \citep[e.g.,][]{Ramos_Almeida17,Hickox18}.
The ionization parameter $\xi$ can be calculated with the ionizing photon flux and the hydrogen number density, and $\xi$ of the BLR is estimated to be $-1\lesssim\log\xi\lesssim1.5$ based on these data in the literature \citep[e.g.][]{Korista04,Lawther18,Dehghanian20}.
In this study, the conservative upper limit of $\xi$ of the dust-free gas, $\log\xi\sim0$, is approximately comparable to that of the BLR.
\cite{Peterson06} mentioned that the BLR must have a large gas reservoir and the total amount of the gas in the BLR is much larger than that expected from the observation of the emission lines.
This reserved gas component may therefore contribute to the dust-free gas obscuration.


One reason for the possible difference in the typical ionization state of the dust-free gas between the obscured and unobscured AGN can be that the column density of the dust-free gas is typically much larger in the obscured AGN. 
Consequently, most of the ionizing photon is absorbed by a small fraction of the gas and a large part of the dust-free gas remains in the neutral or low-ionization state in the obscured AGN.
Another possible reason is that the ionizing flux from the accretion disk is intrinsically small in the Sy1.9 or Sy2 AGN.
This scenario is often discussed in the study of the changing-look (state) AGN because the optical type transition can be associated with the change of the X-ray flux or the AGN bolometric luminosity \citep[e.g.,][]{Noda18,Ricci&Trakhtenbrot22}.

Consequently, our result suggests that AGNs generally have a dust-free gas structure whose spatial scale may be comparable to that of the BLR, and it may have different ionization states depending on the Seyfert type.
This result is similar to that of \cite{Noda23}, wherein they performed Fe-K$\alpha$ reverberation mapping analysis of a changing-look AGN, NGC3516, and showed that the BLR material is retained there during the type-2 phase.

\begin{figure*}
    \includegraphics[width=0.7\linewidth]{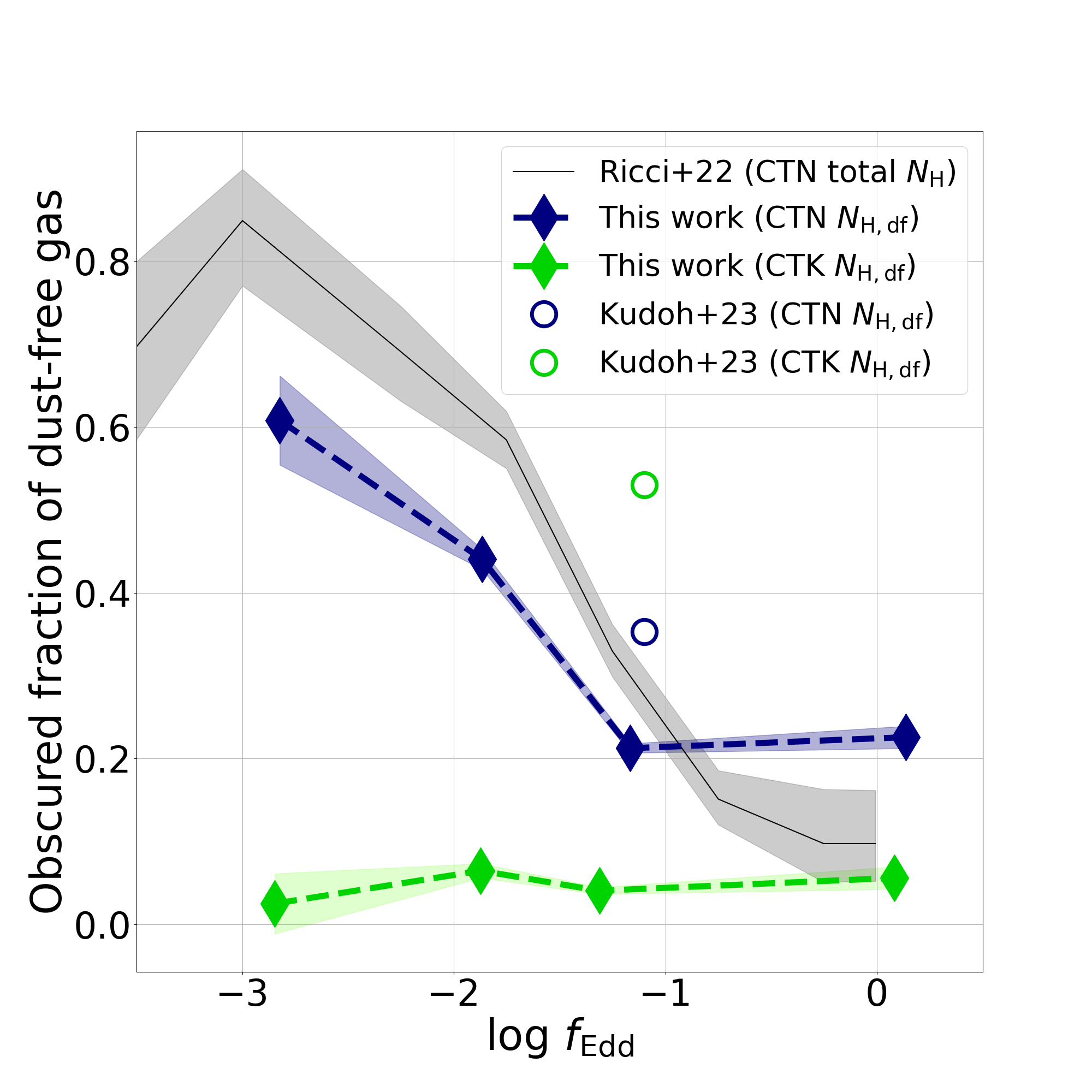}\par 
\caption{The fraction of the target with the dust-free gas column density of $22\leq\log N_{\mathrm{H,df}}\ [\mathrm{cm^{-2}}]<24$ (Compton thin, or CTN, navy plot) and $\log N_{\mathrm{H,df}}\ [\mathrm{cm^{-2}}]\geq24$ (Compton thick, or CTK, green plot). 
The coloured shaded regions along these plots represent $1\sigma$ uncertainty of these fractions.
The black line with gray shaded region indicates the fraction of the target with the total gas column density of $22\leq\log N_{\mathrm{H}}\ [\mathrm{cm^{-2}}]<24$ and its $1\sigma$ uncertainty (Ricci et al. 2022b).
The navy and green open circles indicate the covering factor of the Compton-thin and Compton-thick dust-free gas based on Kudoh et al. (2023), respectively.
}
\label{fig:dust-free-gas covering factor}
\end{figure*}

\subsubsection{dust-free gas covering factor and its Eddington-ratio dependence}
\label{subsubsec:dust-free gas covering factor}

Similar to the discussion in Sec. \ref{subsubsec:dusty gas covering factor}, we calculated the fraction of the AGN in each $N_{\mathrm{H,df}}$ bin of $22\leq\log N_{\mathrm{H,df}}\ [\mathrm{cm^{-2}}]<24$ (Compton-thin dust-free gas) and $\log N_{\mathrm{H,df}}\ [\mathrm{cm^{-2}}]\geq24$ (Compton-thick dust-free gas) as a function of the Eddington ratio.
The method of this calculation follows that presented in Sec. \ref{subsubsec:dusty gas covering factor} and we here set the $1\sigma$ error of $\log N_{\mathrm{H,df}}$ as 0.24 dex, and 0.11 dex for targets with $22\leq\log N_{\mathrm{H,df}}\ [\mathrm{cm^{-2}}]<24$, and $\log N_{\mathrm{H,df}}\ [\mathrm{cm^{-2}}]\geq24$, respectively (Sec.\ref{subsec:dust-free NH separation}).

Figure \ref{fig:dust-free-gas covering factor} shows the resulted fraction of the AGN in each $N_{\mathrm{H,df}}$ bin and its dependence on the Eddington ratio.
The covering factor of the Compton-thin dust-free gas is $f_{\mathrm{C}}\sim0.6$ in $\log f_{\mathrm{Edd}}\sim-3$ and shows a continuous decrement until it becomes $f_{\mathrm{C}}\sim0.2$ in $\log f_{\mathrm{Edd}}\sim-1$, while it seems rather constant in $\log f_{\mathrm{Edd}}>-1$.
This behaviour of the covering factor of the dust-free gas is qualitatively consistent with that of the Compton-thin AGN fraction \citep{Ricci22}.
This result implies that the Eddington-ratio dependence of the covering factor of the Compton-thin gas around the AGN is primarily constrained by that of the dust-free gas.
The covering factor of the Compton-thin dust-free gas is typically smaller than that of the Compton-thin gas \citep[black thin line in Fig.\ref{fig:dust-free-gas covering factor},][]{Ricci22} in $\log f_{\mathrm{Edd}}\lesssim-1$.
This difference may be owing to the contribution of the dust-free gas with $\log N_{\mathrm{H,df}}\ [\mathrm{cm^{-2}}]<22$ with the combination of the dusty gas (see Sec.\ref{subsec:AGN picture} for detail).

Although the decrease in the covering factor in $\log f_{\mathrm{Edd}}\lesssim-1$ may suggest that the Compton-thin dust-free gas will be blown out as the dust-free gas outflow, the effective Eddington limit of the dust-free gas only covers the $f_{\mathrm{Edd}}$ range of $\log f_{\mathrm{Edd}}\gtrsim-1$ in $\log N_{\mathrm{H}}\ [\mathrm{cm^{-2}}]\gtrsim22$, where the covering factor seems rather constant.
This constant covering factor may be owing to the binning effect.
While the uncertainty becomes larger, the covering factor of the Compton-thin dust-free gas shows a good agreement with that of the Compton-thin gas \citep{Ricci22} when we separate the range $\log f_{\mathrm{Edd}}\geq-1.5$ into four bins: $-1.5\leq\log f_{\mathrm{Edd}}<-1$, $-1\leq\log f_{\mathrm{Edd}}<-0.5$, $-0.5\leq\log f_{\mathrm{Edd}}<0$, and $\log f_{\mathrm{Edd}}\geq0$.
We note that, in this detailed binning, the fraction of the AGN with the Compton-thin dust-free gas exhibits a steep increase in $\log f_{\mathrm{Edd}}\geq0$.
This trend can also be seen in the Compton-thin AGN fraction of the original BASS DR2 catalogue.
Another possible reason is the relatively small number of samples with large $f_{\mathrm{Edd}}$. 
The number of the target with $\log f_{\mathrm{Edd}}\gtrsim-1$ is considered to be intrinsically small in all $N_{\mathrm{H}}$ ranges; hence, the target fraction with the Compton-thin dust-free gas does not largely change even if the radiation-driven dust-free gas outflow occurs.
The decrease in the covering factor of the Compton-thin dust-free gas in $\log f_{\mathrm{Edd}}\lesssim-1$ may be partly attributed to the dust-free gas outflow which is not driven by the radiation pressure (Sec.\ref{subsec:dust-free gas outflow} presents further details).

We did not observe a clear Eddington-ratio dependence for the covering factor of the Compton-thick dust-free gas, which is the same trend as the covering factor of the Compton-thick gas in \cite{Ricci17c}.
As we explained in Sec.\ref{sec:intro}, this trend can be explained as follows: if the dust-free gas has a large column density, the ionizing photon will be absorbed by only  the front side of the gas with $N_{\mathrm{H,df}}$ of a few $\times10^{21}\ [\mathrm{cm}^{-2}]$. 
The remaining gas in the back side works as a 'dead weight' \citep{Fabian08}, which prevents the gas from being blown out.
In Fig. \ref{fig:dust-free-gas covering factor}, we show the covering factor of the Compton-thin and Compton-thick dust-free gas based on \cite{Kudoh23}.
As in Fig. \ref{fig:Av_fraction}, these values are much larger than our result.
We also summarized the calculated fraction of the AGN with the Compton-thin and Compton-thick dust-free gas in Tab.\ref{tab:covering factor}.

\begin{table*}
 \caption{Observed dusty/dust-free gas obscured fraction in this study.}
 \label{tab:covering factor}
\centering
 \begin{tabular}{lcccc}
  \hline & $\log f_{\mathrm{Edd}}<-2.5$ & $-2.5\leq\log f_{\mathrm{Edd}}<-1.5$ & $-1.5\leq\log f_{\mathrm{Edd}}<-0.5$ & $\log f_{\mathrm{Edd}}\geq-0.5$\\
  \hline
$A_V\geq5$ mag & $0.60\pm0.08$ & $0.56\pm0.02$ & $0.32\pm0.02$ & $0.36\pm0.05$\\
$A_V\geq40$ mag & $0.07\pm0.04$ & $0.07\pm0.01$ & $0.04\pm0.01$ & $0.01\pm0.01$\\
  \hline
$22\leq\log N_{\mathrm{H,df}}\ [\mathrm{cm^{-2}]}<24$ & $0.61\pm0.05$ & $0.44\pm0.01$ & $0.21\pm0.01$ & $0.23\pm0.01$\\
$\log N_{\mathrm{H,df}}\ [\mathrm{cm^{-2}}]\geq24$ & $0.03\pm0.04$ & $0.06\pm0.01$ & $0.04\pm0.01$ & $0.05\pm0.01$\\

  \hline
\end{tabular}
\end{table*}

\section{discussion}
\label{sec:discussion}

\subsection{The Eddington ratio dependence of the dust-free gas covering factor}
\label{subsec:dust-free gas outflow}

In this study, we show that the covering factor of the Compton-thin dust-free gas decreases in larger Eddington ratio in $-3\lesssim\log f_{\mathrm{Edd}}\lesssim-1$ (see Sec.\ref{subsubsec:dust-free gas covering factor}).
Similar to the decrease of the covering factor of the dusty gas (Sec. \ref{subsubsec:dusty gas covering factor}), this may be owing to the dust-free gas outflow blowing out dust-free gas components more in a larger Eddington ratio.
Certain types of AGN dust-free gas outflow have been observed in the X-ray and UV band with absorption line features \citep[][ for review]{Laha21,Gallo23}.
Here, we compare the expected properties of the dust-free gas outflow with the properties of the warm absorber \citep[][]{Reynolds95,Winter12,Laha14}, which is considered to partly originate in the scale of the dust-free gas region.

The warm absorber is an X-ray absorbing material which is generally observed as an absorption feature or an absorption edge of the H- or He-like ion of, for instance, C, O, and N \citep[e.g.][]{Halpern84,Fabian94,Reynolds97,Costantini00}.
These features typically exhibit blueshifts; hence, the warm absorber is considered to be the ionized gas outflow with the velocity of $v_{\mathrm{out}}\sim10^2-10^3\ \mathrm{km\ s^{-1}}$ \citep[e.g.][]{Laha21,Gallo23}.
Certain studies investigated the properties of the warm absorber in dozens of nearby Sy1 AGNs \citep[e.g.][]{McKernan07,Winter12,Laha14} and found that the warm absorber had a wide range of both the column density ($\log N_{\mathrm{H}}\ [\mathrm{cm^{-2}}]\sim21$--$23$) and the ionization parameter ($\log \xi\sim-1$--$3$).
These studies also showed that the detection rate of the warm absorber in nearby Sy1 AGNs exceeded 50\%.
Based on the AGN unified model, this result indicates that the covering factor of the warm absorber is larger than approximately $0.5$.

\cite{Reynolds95} calculated the typical spatial scale of the warm absorber to be $10^{15-18}\ \mathrm{cm}$, or $\sim0.001-1\ \mathrm{pc}$, from the central SMBH based on the observation of the warm absorber in MCG-6-30-15.
\cite{Blustin05} suggested that the warm absorber mainly originated in the dusty torus.
\cite{Mehdipour18} also indicated the presence of dust in the warm absorber through comparison of the X-ray-measured gas column density and the IR reddening.
However, a recent radiation-hydrodynamic simulation study \citep{Ogawa22} suggested that, although the dusty torus may be the origin of the warm absorber in part, we should consider the outflow component that originates in the region closer to the SMBH than the dusty torus to reproduce the observed absorption features in the X-ray spectrum.

Considering the wide range of the ionization parameter of the warm absorber, it may be a mix of outflow components with various ionization states \citep[e.g.][]{Mehdipour18}.
The conservative upper limit of the ionization parameter of the dust-free gas in Fig \ref{fig:dust-free-gas_Eratio}, that is, $\log\xi\sim0$, is within the range of that of the warm absorber \citep[e.g.][]{McKernan07,Winter12,Laha14,Mehdipour18}.
In Fig.\ref{fig:dust-free-gas_Eratio}, the region with a larger Eddington ratio than that corresponding to the effective Eddington limit of the dust-free gas with $\log\xi=0$ covers the $N_{\mathrm{H}}$ range of $\log N_{\mathrm{H}}\ [\mathrm{cm^{-2}}]\lesssim23$, which may also be similar to that of the warm absorber.
However, we cannot directly compare them because $N_{\mathrm{H}}$ in this study was derived assuming the cold gas \citep{Ricci17b}, as mentioned in Sec.\ref{subsubsec:dust-free gas vs fEdd}.
The high detection rate of the warm absorber in the nearby Sy1 AGNs and the low detection rate in powerful quasars \citep[e.g.][]{Reynolds95} may be related to the drop of the covering factor of the dust-free gas in a high-Eddington regime in Fig.\ref{fig:dust-free-gas covering factor}.
Consequently, the dust-free gas outflow may partly contribute to the low-ionized warm absorber with $\log\xi\gtrsim0$ in the high-Eddington state.

However, although the covering factor of the Compton-thin dust-free gas decreases in the Eddington-ratio range of $\log f_{\mathrm{Edd}}\lesssim-1$ in Fig.\ref{fig:dust-free-gas covering factor}, the effective Eddington limit of dust-free gas only covers the $N_{\mathrm{H,df}}$ range of $\log N_{\mathrm{H,df}}\ [\mathrm{cm^{-2}}]\lesssim22$ when $\log f_{\mathrm{Edd}}\lesssim-1$ in Fig. \ref{fig:dust-free-gas_Eratio}.
This result indicates that it is difficult to explain the decrease in the covering factor of the Compton-thin dust-free gas with the radiation-driven dust-free gas outflow in the low-Eddington state and there are certain other mechanisms that cause the dust-free gas outflow.
\cite{Mizumoto19} assumed the thermal-driven wind as a primary component of the warm absorber and showed that it can be launched from the BLR in low mass-accretion-rate AGNs.
However, they also suggested that the primary component of the wind can be driven not by the thermal pressure but rather by the radiation pressure to the dust.
\cite{Mehdipour19} showed a clear anti-correlation between the column density of the warm absorber and the radio loudness for radio-loud AGNs, which exhibit typically low Eddington ratios \citep[e.g.][]{Fabian12}.
This result suggests the possible contribution of the magnetically-driven outflow in the low-Eddington state. 
Kudoh et al. (in prep) discussed a scenario wherein both the dusty and dust-free gas outflows were affected by inner accretion-disk-scale line-driven winds based on the Eddington-ratio dependence of $N_{\mathrm{H}}$ in their simulations.
Consequently, our results suggest that the radiation-driven dust-free gas outflow may occur in the high-Eddington state of $\log f_{\mathrm{Edd}}\gtrsim-1$, whereas the decrease in the dust-free gas covering factor in $\log f_{\mathrm{Edd}}\lesssim-1$ may be owing to the dust-free gas outflow, which is driven by certain other mechanisms than the radiation pressure.


\begin{figure*}
    \includegraphics[width=\linewidth]{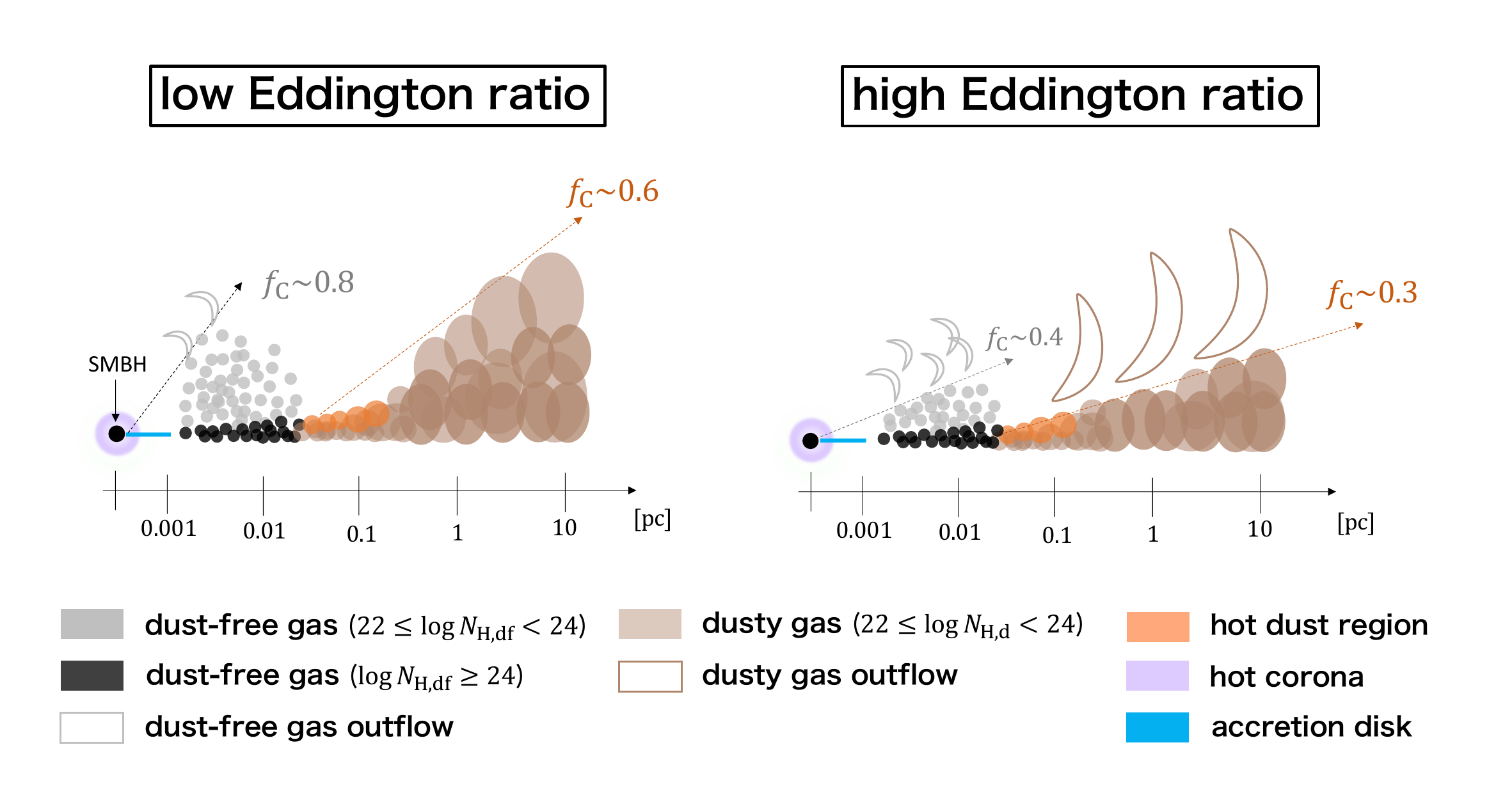}\par 
\caption{The schematic view of the AGN gas structure suggested in this study. The colours of each structure are described in the legend. The arrows indicate the approximate covering factor of the total dusty gas structure (red-brown arrow) and the total dust-free gas (gray arrow), respectively. 
The covering factor of the Compton-thin dusty, dust-free gas structures is based on the result of this study, while that of the Compton-thick dust-free gas structure is based on Ricci et al. (2017b) (see Sec.\ref{subsubsec:Comparison with Ricci+} in detail).}
 \label{fig:picture}
\end{figure*}

\subsection{Updated picture of the AGN gas structures}
\label{subsec:AGN picture}

Figure \ref{fig:picture} shows the geometrical picture of the dusty gas and dust-free gas structures around the low-Eddington and high-Eddington AGN based on the results of this study.
This is a type of updated picture of the AGN gas structure suggested by \cite{Ricci17c}.
Although certain radiation-hydrodynamic simulation studies \citep[e.g.][]{Wada12,Wada16,Kudoh23} suggested that the AGN gas structure be dynamic and partly have time variations, we based our discussion on the typical gas column density for simplicity.
We draw the gas structures in a "fan-like" shape following the schematic images in certain previous studies \citep[e.g.][]{Ricci17c,Honig19}, while certain other studies suggested that these gas structures exhibit a thick-disk shape \citep[e.g.][]{Baskin18}.

\subsubsection{Comparison with Ricci et al. (2017b)}
\label{subsubsec:Comparison with Ricci+}

\cite{Ricci17c} suggested the intrinsic covering factor of the Compton-thin gas structure to be $f_{\mathrm{C}}=0.64\pm0.05$ in the low-Eddington state, and $f_{\mathrm{C}}=0.19\pm0.04$ in the high-Eddington state.

In the low-Eddington state, the covering factor of both the Compton-thin dusty gas and the Compton-thin dust-free gas is similar ($f_{\mathrm{C}}\sim0.6$) and both of them are comparable to that of the Compton-thin gas structure in \cite{Ricci17c}.
However, they are smaller than the Compton-thin AGN fraction in \cite{Ricci22}, as shown in Fig. \ref{fig:dust-free-gas covering factor}.
The possible reason of this smaller covering factor in this study may be owing to the separation of the gas component into the dusty and dust-free parts; in other words, there may be certain solid angles wherein the total gas column density is $22\leq\log N_{\mathrm{H}}\ [\mathrm{cm^{-2}}]<24$, whereas the column density of both the dusty and dust-free gas are $\log N_{\mathrm{H,d/df}}\ [\mathrm{cm^{-2}}]<22$.
In the high-Eddington state, the covering factor of the Compton-thin dust-free gas is consistent with that of the Compton-thin gas \citep{Ricci17c}, although the covering factor of the Compton-thin dusty gas ($f_C\sim0.3$) is slightly larger than it.
This may be owing to the possible overestimation of the dust-obscured fraction in the high-Eddington state (Sec.\ref{subsubsec:dusty gas covering factor}).

The covering factor of the Compton-thick dust-free gas is $f_{\mathrm{C}}\sim0.05$ in this study regardless of the Eddington ratio (Fig.\ref{fig:dust-free-gas covering factor}).
Although this value is smaller than that of the Compton-thick gas structure in \cite{Ricci17c}, the observed target fraction with the Compton-thick gas in BASS DR1 catalogue \citep[$\sim0.075$,][]{Ricci17b} was comparable to the covering factor of the Compton-thick dust-free gas in this study.
\cite{Ricci17c} estimated the intrinsic covering factor of the gas structure from the target fraction by correcting the observational bias of the X-ray absorption.
Therefore, the intrinsic covering factor of the Compton-thick dust-free gas may also be as large as $f_{\mathrm{C}}\sim0.2$.
We adopted this covering factor in Fig. \ref{fig:picture}.

As mentioned in Secs. \ref{subsec:NH distribution} and \ref{subsubsec:dusty gas covering factor}, the covering factor of the Compton-thick dusty gas is expected to be very small.
Although there are certain possibilities wherein the intrinsic covering factor of the Compton-thick dusty gas may be larger than the observation as that of the dust-free gas, $N_{\mathrm{H,d}}$ of the targets with the Compton-thick dust-free gas is entirely Compton thin in this study.
This suggests that, even if the Compton-thick dusty gas is actually present, its covering factor is smaller than that of the Compton-thick dust-free gas.
Therefore, we did not explicitly show the Compton-thick dusty gas structure in Fig. \ref{fig:picture}.


\subsubsection{Dusty gas structure}
\label{subsubsec:dusty gas structure}

Most fraction of the dusty gas with $A_V\geq5$ mag is considered to be distributed in the dusty torus.
The spatial scale of the dusty torus has been investigated with the dust reverberation mapping \citep{Suganuma06,Koshida14,Minezaki19,Lyu19,Yang20,Noda20} and its inner edge is approximately $0.1$ pc.
These studies found that the reverberation radius depends on the AGN luminosity as $r\propto L^{1/2}$ and showed that the inner edge of the dusty torus is likely to be constrained by the dust sublimation, and such innermost region is considered to comprise the hot dust with a temperature of $T\sim1000$--$1500$ K.
Furthermore, recent observations with the MIR interferometer \citep[e.g.][]{Gravity20,Gamez_Rosas22} or Atacama Large Millimeter/sub-millimeter Array \citep[ALMA, e.g.][]{Impellizzeri19, Garcia-Burillo19,Garcia-Burillo21,Izumi23} has facilitated the application of a direct constrain to the size of the dusty torus to be smaller than tens of parsecs. 

In the low-Eddington state, the elevation angle $\theta$ ($\theta=0^{\circ}$ indicates the edge-on view) of the dusty gas structure, that is, the dusty torus, is estimated to be approximately $\theta\sim37^{\circ}$, or the covering factor of $f_{\mathrm{C}}\sim0.6$ in this study.
The majority portion of the dusty torus is expected to be Compton thin.
The dusty gas structure is not likely to have the powerful dusty gas outflow in the low-Eddington state, or $\log f_{\mathrm{Edd}}\lesssim-2$; thus, it is relatively stable when the AGN does not have a powerful accretion.

However, in the high-Eddington state, $\theta$ of the dusty torus decreases to be $\theta\sim15^{\circ}$, or the covering factor of $f_{\mathrm{C}}\sim0.3$.
This is primarily owing to the radiation-driven dusty gas outflow blowing out certain fraction of the dusty gas.
This blown-out dusty gas may contribute to the polar dust component \citep[e.g.][]{Honig12,Burtscher13,Tristram14,Lopez-Gonzaga16,Leftley18,Isbell22,Ogawa21}.
The typical dust extinction of the polar dust is excessively small to measure with the method of \cite{Mizukoshi22} \citep[$\lesssim1$ mag,][]{Buat21}.
Although a possible compact hot polar dust structure  \citep[][and citation therein]{Yamada23} may have a larger dust extinction, we cannot distinguish it from the dust extinction owing to the dusty torus.
Therefore, we did not discuss the structure of the polar dust in detail.
We note that a certain fraction of the dusty gas outflow may be observed as the warm absorber in the very-low ionization state that contains a certain amount of dust \citep[e.g.][]{Blustin05,Mizumoto19}.

\subsubsection{Dust-free gas structure}
\label{subsubsec:dust-free gas structure}

As explained in Sec.\ref{subsubsec:dust-free gas vs fEdd}, the dust-free gas is expected to be in the same scale as the BLR.
The ionization state of the dust-free gas may also be consistent with this scenario, as hinted by Fig. \ref{fig:dust-free-gas_Eratio}. 
\cite{Minezaki19} compared the \textit{K}-band dust reverberation radius and the broad H$\beta$ reverberation radius \citep{Bentz13} and showed that the broad H$\beta$ emitting region is approximately $0.6$ dex smaller than the inner edge of the dusty torus.
Furthermore, \cite{Homayouni23} performed reverberation mapping measurements for several ultraviolet broad lines of Mrk 817 and showed the typical time scale of the reverberation lag to be $\sim10$ days, which is approximately $30$--$50\%$ of that of the broad H$\beta$ line of the same object \citep{Bentz13}.
Considering that the dust-free gas may exist primarily in the BLR, the dust-free gas is also considered to cover the range of approximately $0.01$--$0.1$ pc from the central source, while it depends on the SMBH mass.

For $\log f_{\mathrm{Edd}}\sim-3$, $\theta$ of the entire dust-free gas structure is estimated to be $\theta\sim53^{\circ}$, or the covering factor of $f_{\mathrm{C}}\sim0.8$ in this study.
The covering factor of the Compton-thin dust-free gas is $f_{\mathrm{C}}\sim0.6$, whereas we set the covering factor of the Compton-thick dust-free gas as $f_{\mathrm{C}}\sim0.2$ based on \cite{Ricci17c} (Sec.\ref{subsubsec:Comparison with Ricci+}).
For $\log f_{\mathrm{Edd}}\sim-1$, $\theta$ of the entire dust-free gas structure decreases to be $\theta\sim25^{\circ}$, or the covering factor of $f_{\mathrm{C}}\sim0.4$.
Here, unlike the case of the dusty gas, the effect of the radiation-driven dust-free gas outflow is considered to be nearly absent because the effective Eddington limit of the dust-free gas is $\log f_{\mathrm{Edd}}\gtrsim-1$ in $\log N_{\mathrm{H,df}}\ [\mathrm{cm^{-2}}]>22$.
However, the dust-free gas outflow driven by the magnetic structure \citep[e.g.][]{Mehdipour19} or the accretion-disk-scale winds (e.g. Kudoh et al. in prep) may have certain effects on the dust-free gas structure for $\log f_{\mathrm{Edd}}\lesssim-1$.
Similar to the result of \cite{Ricci17c}, the covering factor of the Compton-thick dust-free gas does not significantly change regardless of the Eddington ratio (Sec.\ref{subsec:dust-free gas}).

\section{Conclusion}
\label{sec:conclusion}

This study derived the line-of-sight dust extinction $A_V$ of 589 nearby X-ray selected AGNs in BASS DR2 catalogue.
By combining this dust extinction and the neutral gas column density $N_{\mathrm{H}}$, we separated the dusty and dust-free gas components of the AGN gas structure and investigated their physical properties and relations to the radiation pressure.
Our primary findings are as follows:

\begin{enumerate}[1.]
   \item The typical column density of the dusty gas ($N_{\mathrm{H,d}}$) and the dust-free gas ($N_{\mathrm{H,df}}$) is different in each Seyfert type. 
   This result can be explained based on the orientation effect, wherein both the dusty and dust-free gas structures have an anisotropic column density distribution and the observed column density varies depending on the viewing angle.\\
   
   \item The total distributions of $N_{\mathrm{H,d}}$ and $N_{\mathrm{H,df}}$ differs and there may be very few targets with $\log N_{\mathrm{H,d}}\ [\mathrm{cm^{-2}}]>23$, while a certain number of targets show $\log N_{\mathrm{H,df}}\ [\mathrm{cm^{-2}}]>23$.
   Although this result may imply that the covering factor of the heavily obscuring dusty gas is very small compared to that of the dust-free gas with a similar column density, the effect of the observational bias cannot be excluded. \\
   
   \item In the comparison between $A_V$ and the Eddington ratio $f_{\mathrm{Edd}}$, very few targets are distributed in the forbidden region, which is consistent with many previous studies.
   By using $A_V$ instead of $N_{\mathrm{H}}$, which is commonly used in the literature, we can neglect the possible effects of the dust-free gas and then clearly show many obscured AGNs being distributed concentrically just out of or on the boundary of the forbidden region.\\
   
   \item The covering factor of the dusty gas with $A_V\geq5$ mag demonstrated a clear drop at $-2\lesssim\log f_{\mathrm{Edd}}\lesssim-1$ and otherwise it is relatively constant.
   This $f_{\mathrm{Edd}}$ range is consistent with the effective Eddington limit of the Compton-thin dusty gas.
   Therefore, this result supports the scenario of the literature, wherein the covering factor of the dusty gas decreases in a high Eddington ratio owing to the radiation-driven dusty gas outflow.
   The covering factor of the dusty gas with $A_V\geq40$ mag does not significantly change regardless of the Eddington ratio.\\


   \item Similar to the dusty gas structure, the covering factor of the Compton-thin dust-free gas also decreases with the increase of the Eddington ratio.
   Although this decrease may also be owing to the dust-free gas outflow, this outflow is less likely to be owing to the radiation pressure in $\log f_{\mathrm{Edd}}\lesssim-1$, because the effective Eddington limit of the partially-ionized dust-free gas corresponds to $\log f_{\mathrm{Edd}}\gtrsim-1$ in the Compton-thin regime. 
   Such dust-free gas outflow may be driven by the magnetic structure or the accretion-disk-scale line-driven winds in the low-Eddington state.
   Certain fraction of the dust-free gas outflow may contribute to the ionized gas outflow such as the warm absorber.
   Similar to the dense dusty gas structure, the covering factor of the Compton-thick dust-free gas is relatively constant regardless of the Eddington ratio.\\

   \item This study proposed the schematic picture of the gas structure of the AGN in both the low- and high-Eddington states, which is a type of the update of \cite{Ricci17c}.
   The dusty gas is distributed within $\sim1$--10 pc from the centre of the AGN.
   Further, the covering factor of the dusty gas is approximately $f_{\mathrm{C}}\sim0.6$ in the low-Eddington state and it decreases to $f_{\mathrm{C}}\sim0.3$ in the high-Eddington state owing to the radiation-driven dusty gas outflow.
   The dust-free gas structure is considered to be distributed at the same scale as that of the BLR ($\sim0.01$--0.1 pc) based on the time scale of the $N_{\mathrm{H}}$ variation \citep[e.g.][]{Burtscher16}, radiation-hydrodynamic simulation \citep{Kudoh23}, and the rough estimation of the ionization state of the dust-free gas.
   The covering factor of the Compton-thin dust-free gas is approximately $f_{\mathrm{C}}\sim0.6$ in the low-Eddington state and it decreases to $f_{\mathrm{C}}\sim0.2$ in the high-Eddington state; further, we adopt $f_{\mathrm{C}}\sim0.2$ from \cite{Ricci17c} as the covering factor of the Compton-thick dust-free gas in both low- and high-Eddington states.

\end{enumerate}

\section*{Acknowledgements}
We thank the anonymous referee for many constructive comments which improve the quality of this paper.
We also thank Kohei Ichikawa, Yoshihiro Ueda, Keiichi Wada, Yuki Kudoh, and Claudio Ricci for their many insightful comments.
S.M. is thankful for financial supports by Japan Society for the Promotion of Science (JSPS) Research Fellowship for Young Scientists (Nos.23KJ0450).
S.M. is also supported by JST SPRING, Grant number JPMJSP2108.
H.N. is supported by JSPS KAKENHI Grant number 19K21884, 20KK0071, 20H0941, and 20H01947.
H.S. is supported by JSPS KAKENHI Grant number 19K03917.
T.K. is grateful for support from RIKEN Special Postdoctoral Researcher Program and is supported by JSPS KAKENHI Grant number JP23K13153.

This publication has made use of data products from the \it{}Wide-field Infrared Survey Explorer\rm{}, which is a joint project of the University of California, Los Angeles, and the Jet Propulsion Laboratory/California Institute of Technology, funded by the National Aeronautics and Space Administration. This publication also makes use of data products from \it{}NEOWISE\rm{}, which is a project of the Jet Propulsion Laboratory/California Institute of Technology, funded by the Planetary Science Division of the National Aeronautics and Space Administration.

We also acknowledge the use of public data from the BAT AGN Spectroscopic Survey.

This research has made use of the NASA/IPAC Extragalactic Database (NED), which is operated by the Jet Propulsion Laboratory, California Institute of Technology, under contract with the National Aeronautics and Space Administration.

\section*{data availability}
The \it{}WISE\rm{} data used in this study are publicly available in the NASA/IPAC Infrared Science Archive (\url{https://irsa.ipac.caltech.edu/Missions/wise.html}). 
The data of BASS AGN catalogue used in this study are also publicly available from the BASS website (\url{https://www.bass-survey.com}).
All data derived in this study are available in the online version of this paper.




\bibliographystyle{mnras}
\bibliography{WISE_1} 







\bsp	
\label{lastpage}
\end{document}